\documentclass[aps,prd,onecolumn,groupedaddress,showpacs,nofootinbib,amssymb
]{revtex4}
\usepackage[dvips]{graphicx}
\usepackage{amssymb}
\usepackage{amsmath}
\usepackage{epstopdf}
\usepackage{graphicx,,color}
\usepackage{amsfonts}
\usepackage{bm}
\usepackage{cancel}
\usepackage{comment}

\allowdisplaybreaks[4]

\begin{document}

\tolerance=5000



\title{Constraints on prospective deviations from the cold dark matter model using a Gaussian Process}

\author{
Martiros Khurshudyan\thanks{Email: khurshudyan@ice.csic.es} and
Emilio Elizalde\thanks{E-mail: elizalde@ice.csic.es}}

\affiliation{
Institut of Space Sciences, ICE/CSIC-IEEC,
Campus UAB, Carrer de Can Magrans s/n, 08193 Bellaterra (Barcelona) Spain \\}

\begin{abstract}

Recently, using Bayesian Machine Learning, a deviation from the cold dark matter model on cosmological scales has been put forward. Such model might replace a proposed non-gravitational interaction between dark energy and dark matter, and help solve the $H_{0}$ tension problem. The idea behind the learning procedure relied there on a generated expansion rate, while the real expansion rate was just used to validate the learned results. In the present work, however, the emphasis is put on a Gaussian Process (GP) with the available $H(z)$ data confirming the possible existence of the already learned deviation. Three cosmological scenarios are considered: a simple one, with equation of state parameter for dark matter $\omega_{dm} = \omega_{0} \neq 0$, and two other models, with corresponding parameters  $\omega_{dm} = \omega_{0} + \omega_{1} z$ and  $\omega_{dm} = \omega_{0} + \omega_{1} z/(1+z)$. The constraints obtained on the free parameters $\omega_{0}$ and $\omega_{1}$ hint towards a dynamical nature of the deviation. The dark energy dynamics is also reconstructed, revealing interesting aspects connected with the $H_{0}$ tension problem. It is concluded, however, that improved tools and more data are needed, in order to reach a better understanding of the reported deviation. 

\end{abstract}

\pacs {}

\maketitle

\section{Introduction}\label{sec:INT}

The $H_{0}$ tension problem is presently a hot topic in cosmology. The problem is to understand why the Planck CMB data analysis and a local measurement from the Hubble Space Telescope give different values for $H_{0}$. More specifically, it must be understood why in the $\Lambda$CDM scenario the Planck CMB data analysis gives $H_{0} = 67.4 \pm 0.5$ km/s/Mpc, while local measurements from the Hubble Space Telescope yield $H_{0} = 73.52 \pm 1.62$ km/s/Mpc ~\cite{Aghanim_H0} - \cite{Freedman_H0}. The various mechanisms proposed to explain this unexpected discrepancy need be understood, yet. There is even a hint that the $\Lambda$CDM model could be challenged (see \cite{Perivolaropoulos_LCDM} and \cite{Elizalde_H0} and references therein for more discussion). However, one should stress that, at this stage of research, it is not clear yet if there is a need for new physics or if everything comes from an overseen mistake on the measurement side. 

Anyway, an attempt to challenge the $\Lambda$CDM model was actually in place long before the $H_{0}$ tension problem appeared. Recall, in particular,  the theoretical and conceptual problems caused by the cosmological constant issue, requiring either a modification of general relativity or a consideration of dynamical dark energy models. Indeed, crafting modified theories of gravity and dynamical dark energy models has a relatively long and successful history and plays a crucial role in modern cosmology. In both cases, a workable model can be designed in various ways. For instance, by considering non-gravitational interactions. Alternatively, one can avoid them and use instead a dynamical equation of state parameter for dark energy. On top of that, a non-gravitational interaction allows to alleviate or even solve various problems, and, recently, it has been demonstrated that the $H_{0}$ tension problem may be put among them. Besides all this, reasonable hope is coming from observational data, but from where exactly is it coming is not clear, yet \cite{Perivolaropoulos_LCDM} - \cite{INEnd} (to mention some references covering several aspects of the above-mentioned topics). 

Not to forget, the  $\Lambda$CDM model has a second ingredient, namely cold dark matter, which could be also a source of tension.  But, why the cold dark matter paradigm should be challenged?  To develop some sort of logic or intuition about this issue,  one needs to go back to interacting dark energy models, described by (see for instance \cite{Elizalde_4})
\begin{equation}\label{eq:IntDM0}
\dot{\rho}_{dm} + 3 H (\rho_{dm} + P_{dm}) = Q,
\end{equation}  
and
\begin{equation}\label{eq:IntDE0}
\dot{\rho}_{de} + 3 H (\rho_{de} + P_{de}) = - Q,
\end{equation}
 where $\rho_{de}$ and $\rho_{dm}$ are the dark energy and dark matter energy densities, $P_{de}$ and $P_{dm}$ their pressures, while $Q$ stands for a non-gravitational interaction. The last indicates energy flow between dark energy and dark matter, meaning that their evolution is interconnected and affects each other's dynamics. We can rewrite Eqs. (\ref{eq:IntDM0}) and (\ref{eq:IntDE0}) in a slightly different way, as
\begin{equation}\label{eq:IntDM}
\dot{\rho}_{dm} + 3 H \rho_{dm} \left (1 + \frac{P_{dm}} {\rho_{dm}}- \frac{Q}{\rho_{dm}} \right ) = 0
\end{equation}  
and
\begin{equation}\label{eq:IntDE}
\dot{\rho}_{de} + 3 H \rho_{de} \left (1 + \frac{P_{de}} {\rho_{de}} + \frac{Q}{\rho_{de}} \right ) = 0,
\end{equation} 
where $\omega_{dm} = P_{dm}/\rho_{dm}$ and $\omega_{de} = P_{de}/\rho_{de}$ are the equation of state parameters for dark matter and dark energy, respectively. In this way, we have introduced some effective energy sources where the equation of state parameter depends on $Q$, too. Therefore, even if  $\omega_{dm} = 0$, we already have an effective non-cold dark matter involved in the dynamics. Moreover,  since observational data support $Q \neq 0$, from here it follows that, by introducing a non-gravitational interaction, we induce corrections to both dark energy and dark matter, which are not in contradiction with recent observational data. Now, from the above discussion, it seems quite natural to search for possible deviations from the cold dark matter paradigm. It is interesting to study how this approach could solve, or at least alleviate the already-mentioned tension problems. In particular, it is worthy to discuss the possible impact of the mentioned deviation on the $H_{0}$ tension problem. Understanding this point is crucial for present-day cosmology since we have seen that this deviation could replace the non-gravitational interaction between these two dark sources, which do not share the same origin and do not operate on the same scales. 

Recently, motivated by this possibility, clues for a deviation from the cold dark matter scenario have been gained from a learning procedure, which allowed us to advance towards the solution of the $H_{0}$ tension problem \cite{Elizalde_H0}. The learning method was based on a Bayesian (Probabilistic) Machine Learning approach and thereby generated expansion rates. In Bayesian Machine Learning, one uses a model-based data generation process to connect physics and our initial belief, to do the learning. This is a very convenient learning approach, which allows one to avoid data-related issues and limitations, and uses instead available data, only, to validate the learned results. We need to stress that the choice of the initial belief can be inferred from the data, too, at the very initial step, before the learning starts, which gives an idea about ranges where the model parameters could be defined. An alternative option is to use the model itself, supplying some reasonable values to the parameters. In short, we have two different ways to start the learning process, but in any case, the final result should be always validated using real observational data. But this can originate some doubts about the learned results; therefore, it is required --and even constitutes an urgent task-- to use more traditional, frequently used tools to first, validate and, second, better understand the learned results. 

In the case at hand, we need to study and validate the deviation from the cold dark matter paradigm and understand how this affects the $H_{0}$ tension problem. To achieve our goal, we use a GP, which is one of the most frequently employed tools in modern cosmology, and it is among the most robust machine learning algorithms available. 

It is an approach requiring data to be involved, which here will consist of $30$-point samples of $H(z)$ deduced from the differential age method and $10$ data-points from the BAO~(see Table~\ref{tab:Table0}). Recently this particular tool was used to obtain a model-independent reconstruction of $f(T)$ gravity \cite{Elizalde_0}. It has also been employed to study the string swampland criteria in the dark energy-dominated universe, in the case of general relativity and $f(R)$ theory of gravity, in \cite{Elizalde_1} and \cite{Elizalde_2}, respectively. Moreover, it has been used in a scalar field potential reconstruction allowing to constrain any given model revealing its connection to swampland, among other \cite{Elizalde_3}. In the literature, there are several important applications of this method, to study cosmological and astrophysical problems (see, \cite{GP_0} - \cite{GP_7}, to mention a few). 

In other words, the goal here is to learn and validate a previously learned deviation from the cold dark matter paradigm, using a GP and available expansion rate data (see Table \ref{tab:Table0}). The goal is to reveal possible dynamics in this deviation. Therefore, besides the model with $\omega_{dm} = \omega_{0} \neq 0$, we have also considered two more models, where the equation of state parameter is given by $\omega_{dm} = \omega_{0} + \omega_{1} z$ and  $\omega_{dm} = \omega_{0} + \omega_{1} z/(1+z)$, respectively. Indeed, the constraints on the free $\omega_{0}$ and $\omega_{1}$ parameters that we obtain reveal a hint about the dynamical nature of this deviation. On the other hand, using GP, we will reconstruct the dark energy equation of the state parameter for each of the models considered. The reconstruction allows to reveal various interesting aspects of the dark energy dynamics and hints towards the conclusion that early dark energy could indeed solve the $H_{0}$ tension problem, among other conclusions.

\begin{table}[t]
  \centering
    \begin{tabular}{ |  l   l   l  |  l   l  l  | p{2cm} |}
    \hline
$z$ & $H(z)$ & $\sigma_{H}$ & $z$ & $H(z)$ & $\sigma_{H}$ \\
      \hline
$0.070$ & $69$ & $19.6$ & $0.4783$ & $80.9$ & $9$ \\
         
$0.090$ & $69$ & $12$ & $0.480$ & $97$ & $62$ \\
    
$0.120$ & $68.6$ & $26.2$ &  $0.593$ & $104$ & $13$  \\
 
$0.170$ & $83$ & $8$ & $0.680$ & $92$ & $8$  \\
      
$0.179$ & $75$ & $4$ &  $0.781$ & $105$ & $12$ \\
       
$0.199$ & $75$ & $5$ &  $0.875$ & $125$ & $17$ \\
     
$0.200$ & $72.9$ & $29.6$ &  $0.880$ & $90$ & $40$ \\
     
$0.270$ & $77$ & $14$ &  $0.900$ & $117$ & $23$ \\
       
$0.280$ & $88.8$ & $36.6$ &  $1.037$ & $154$ & $20$ \\
      
$0.352$ & $83$ & $14$ & $1.300$ & $168$ & $17$ \\
       
$0.3802$ & $83$ & $13.5$ &  $1.363$ & $160$ & $33.6$ \\
      
$0.400$ & $95$ & $17$ & $1.4307$ & $177$ & $18$ \\

$0.4004$ & $77$ & $10.2$ & $1.530$ & $140$ & $14$ \\
     
$0.4247$ & $87.1$ & $11.1$ & $1.750$ & $202$ & $40$ \\
     
$0.44497$ & $92.8$ & $12.9$ & $1.965$ & $186.5$ & $50.4$ \\

$$ & $$ & $$ & $$ & $$ & $$\\ 

$0.24$ & $79.69$ & $2.65$ & $0.60$ & $87.9$ & $6.1$ \\
$0.35$ & $84.4$ & $7$ &  $0.73$ & $97.3$ & $7.0$ \\
$0.43$ & $86.45$ & $3.68$ &  $2.30$ & $224$ & $8$ \\
$0.44$ & $82.6$ & $7.8$ &  $2.34$ & $222$ & $7$ \\
$0.57$ & $92.4$ & $4.5$ &  $2.36$ & $226$ & $8$ \\ 
          \hline
    \end{tabular}
    \vspace{5mm}
\caption{$H(z)$ and its uncertainty $\sigma_{H}$  in  units of km s$^{-1}$ Mpc$^{-1}$. The upper panel consists of thirty samples deduced from the differential age method. The lower panel corresponds to ten samples obtained from the radial BAO method. The table is according to \cite{HTable_0} - \cite{HTable_13}.}
  \label{tab:Table0}
\end{table}
  
The present paper is organized as follows. In section~\ref{sec:DGP} we give some details about GPs indicating the strategy we have used to investigate a deviation from the cold dark matter model, and confirming the recently obtained results where Bayesian Machine Learning was used. The description of the models and the discussion of the results obtained conform section~\ref{sec:NewMod}. In three subsections thereof, the interpretation of the results in each case is explained at large. Finally, a summarized discussion of the key results of our study, together with some conclusions one can draw from them, are to be found in section~\ref{sec:Discussion}.

\section{Gaussian Processes}\label{sec:DGP}

In this section, a brief presentation of the approach used in the paper is done, to gain some intuition about the procedure. The mean, $\mu(x)$, and the two-point covariance function, $K(x,x^{\prime})$, are the key ingredients of a GP, intending to get a continuous realization of
\begin{equation}
\xi(x) \propto GP(\mu(x), K(x,x^{\prime}))
\end{equation}
and its related uncertainty $\Delta \xi(x)$; what allows to build the realization region, described by $\xi(x) \pm \Delta \xi(x)$. The last one is the posterior, which will be formed through a Bayesian iterative process. It allows for the reconstruction of the function representing the data, in a completely model-independent way, directly from the data. The assumption imposing the data error distribution to be Gaussian is a very important aspect of this method. In other words, we need to consider that the observational data is also a realization of the GP.  However, here, the kernel should be chosen manually based on the underlying physics producing the given data. 

In the recent literature, the usefulness of various kernels in addressing different issues in cosmology has been assessed. Some of the kernels most widely used for this purpose are the squared exponent
\begin{equation}\label{eq:kernel1}
K(x,x^{\prime}) = \sigma^{2}_{f}\exp\left(-\frac{(x-x^{\prime})^{2}}{2l^{2}} \right),
\end{equation}
the Cauchy kernel
\begin{equation}\label{eq:kernel3}
K_{C}(x,x^{\prime}) = \sigma^{2}_{f} \left [  \frac{l}{ (x - x^{\prime} )^{2}} + l^{2}\right ].
\end{equation}
and the Matern ($\nu = 9/2$)  kernel
$$K_{M}(x,x^{\prime}) = \sigma^{2}_{f} \exp \left(-\frac{3|x-x^{\prime}|}{l} \right) $$
\begin{equation}\label{eq:kernel2}
\times \left[ 1+ \frac{3 |x-x^{\prime}|}{l} + \frac{27(x-x^{\prime})}{7l^{2}} + \frac{18|x-x^{\prime}|^{3}}{7l^{3}} + \frac{27 (x-x^{\prime})^{4}}{35 l^{4}}\right],
\end{equation}

The $\sigma_{f}$ and $l$ parameters appearing in Eqs. (\ref{eq:kernel1}), (\ref{eq:kernel3}) and  (\ref{eq:kernel2}) are called the hyperparameters. The $l$ parameter represents the correlation length along which the successive $\xi(x)$ values are correlated, while the $\sigma_{f}$ parameter is used to control the variation in $\xi(x)$ relative to the mean of the process. The most suitable values of the hyperparameters are determined from the minimization of the GP marginal likelihood. This means that we need to search for the values of the parameters that maximize the probability that the GP generates the considered data. Similarly to other tools, the GPs have advantages and disadvantages that one should not dismiss. As for other Machine Learning (ML) techniques, a GP learns an underlying latent distribution in the data at hand which strongly affects the learning process, too. Therefore, it is not excluded that the algorithm can fail when an unforeseen situation appears. The goal of GP is to infer the features and not visit every single data point to understand why it is there. Because of this, it is possible that it can fail in doing certain tasks or that it will not be able to deal with unforeseen situations, giving sometimes less confident predictions. This aspect should be treated very seriously to avoid misleading results and bad conclusions and consequences. Fortunately, the size of data used in the case of cosmological applications allows for a sensible reduction of the kernel numbers to be considered (as compared to other situations) and also allows us to perform some quick numerical experiments to find where the problem can hide. Readers are invited to check the following references, dealing with how GPs can be used in cosmology, and to learn the necessary mathematics behind this approach \cite{GP_0} - \cite{GP_7}.

A short note on the notation to be used~(with $8 \pi G = c = 1$) . For the metric 
\begin{equation}
ds^{2} = -dt^{2} + a(t)^{2} \sum_{i =1}^{3} (dx^{i})^{2},
\end{equation}
and Hubble rate 
\begin{equation}\label{eq:F1}
H^{2} = \frac{1}{3}(\rho_{de} + \rho_{dm} ),
\end{equation}
describing the background dynamics of the universe. The energy conservation law for two energy components reads 
\begin{equation}\label{eq:de}
\dot{\rho}_{de} + 3 H (\rho_{de} + P_{de}) = 0,
\end{equation}
and 
\begin{equation}\label{eq:dm}
\dot{\rho}_{dm} + 3 H (\rho_{dm} + P_{dm}) = 0.
\end{equation}
In the above, $P_{de}$ and $\rho_{de}$ are the dark energy pressure and energy density, while $P_{dm}$ and $\rho_{dm}$ are the dark matter pressure and energy density, respectively.   

In this paper, we take $P_{dm} = \omega_{dm} \rho_{dm}$, which together with Eq. (\ref{eq:dm}) allows to determine the dark matter energy density dynamics. The last one allows determining the $\rho_{de}$ from Eq. (\ref{eq:F1}). On the other hand, Eq. (\ref{eq:de}) allows to reconstruct the dark energy pressure, according to 
\begin{equation}\label{eq:Pde}
P_{de} =  - \rho_{de} + \frac{1+z}{3} \rho^{\prime}_{de},
\end{equation}
where the prime stands for the derivative with respect to the redshift $z$. The last equation with $\rho_{de}$ allows to reconstruct the dark energy dynamics. In the next section, we will see that, for the three models considered in this paper, only the reconstruction of $H(z)$ and $H^{\prime}(z)$ is required. This will be done by using the squared exponential, Eq. (\ref{eq:kernel1}), the Cauchy kernel Eq. (\ref{eq:kernel3}), and the Matern kernel ($\nu = 9/2$), Eq. (\ref{eq:kernel2}). The results will hint to: 1) a deviation from the cold dark matter model, 2) to the dynamical nature of this deviation, and 3) to an understanding of how this is connected to the $H_{0}$ tension problem. The details of the strategy being used here will be presented in the next section. As in previous studies, we have used the GaPP~(Gaussian Processes in Python) package developed by Seikel et al \cite{Seikel}.

\section{Models and Results}\label{sec:NewMod}

In this section, we present and discuss the models to be used and the constraints obtained with them. We start with the model with $\omega_{dm} = \omega_{0}$, where $\omega_{0}$ is the free parameter to constrain. During the reconstruction, we use the three kernels given by Eqs. (\ref{eq:kernel1}), (\ref{eq:kernel3}), and (\ref{eq:kernel2}), to understand their differentiated role in the possible deviation from the cold dark matter model. Moreover, with each kernel, we perform the reconstruction assuming three different situations. As a first case, we have allowed the GP to estimate the value of $H_{0}$ based on the available expansion rate data, and have used it to reconstruct $H(z)$ and $H^{\prime}(z)$. In the second case, we have set $H_{0} = 73.52 \pm 1.62$ km/s/Mpc together with the expansion rate data and then performed the reconstruction. Finally, in the last case, we took $H_{0} = 67.4 \pm 0.5$ km/s/Mpc followed by the Planck CMB data analysis and, using it with the expansion rate data, we have reconstructed $H(z)$ and $H^{\prime}(z)$. This strategy allowed to gain more intuition about how the deviation from the cold dark matter model can solve the $H_{0}$ tension problem. To save space, we refer to the upper part of Table~\ref{tab:Table1} to find more about the estimated $H_{0}$ value for the kernels considered, when just the data given in Table~\ref{tab:Table0} has been used in the reconstruction.

\begin{figure}[h!]
 \begin{center}$
 \begin{array}{cccc}
\includegraphics[width=165 mm]{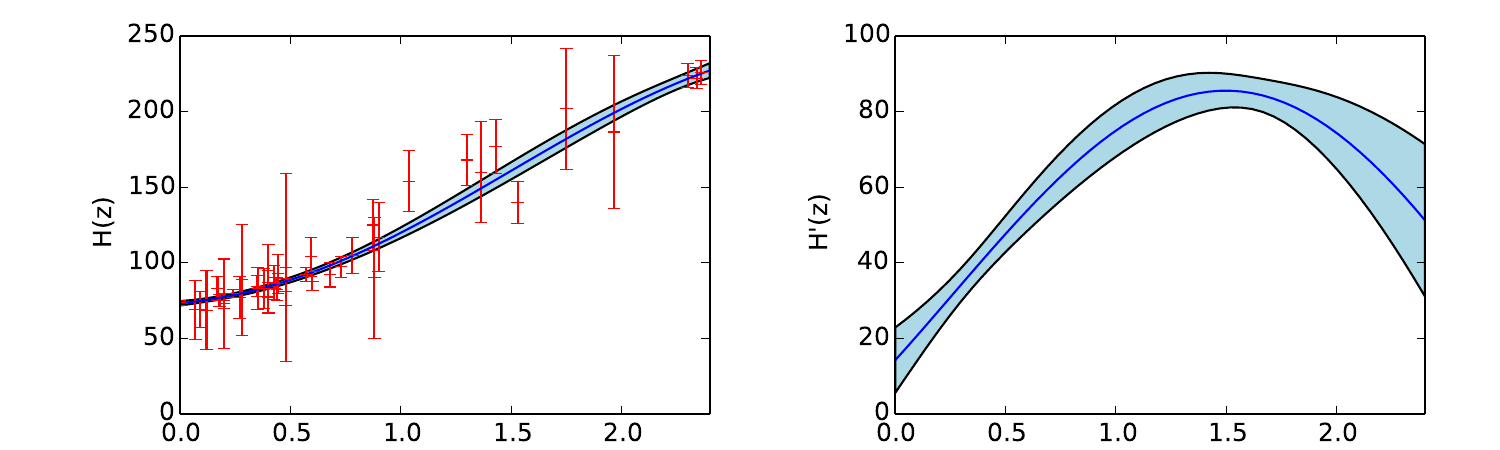} \\
\includegraphics[width=165 mm]{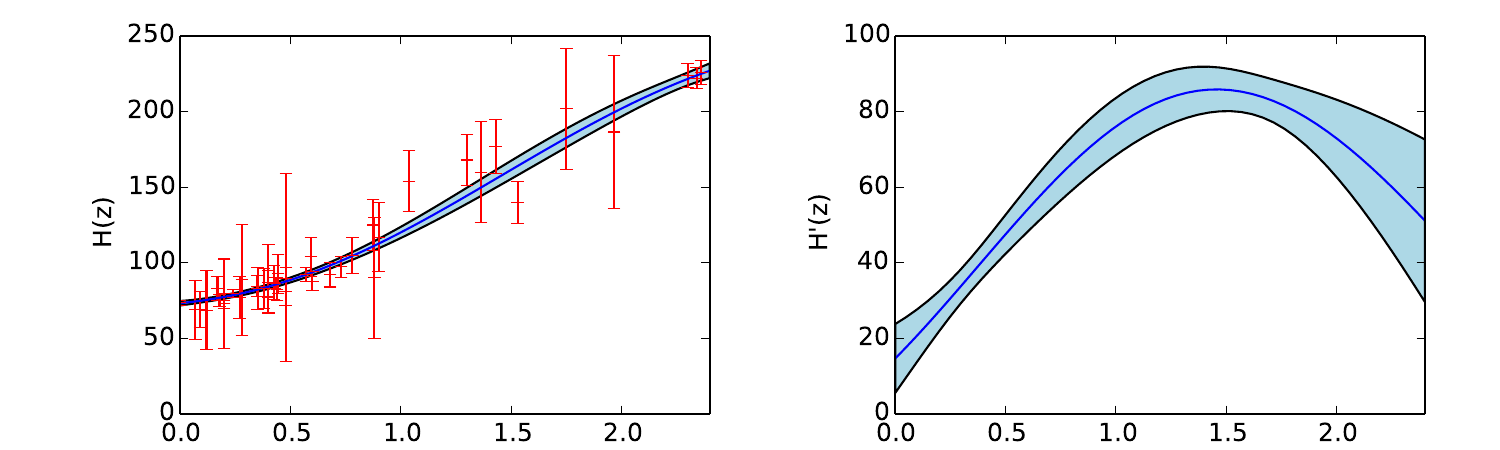} \\
\includegraphics[width=165 mm]{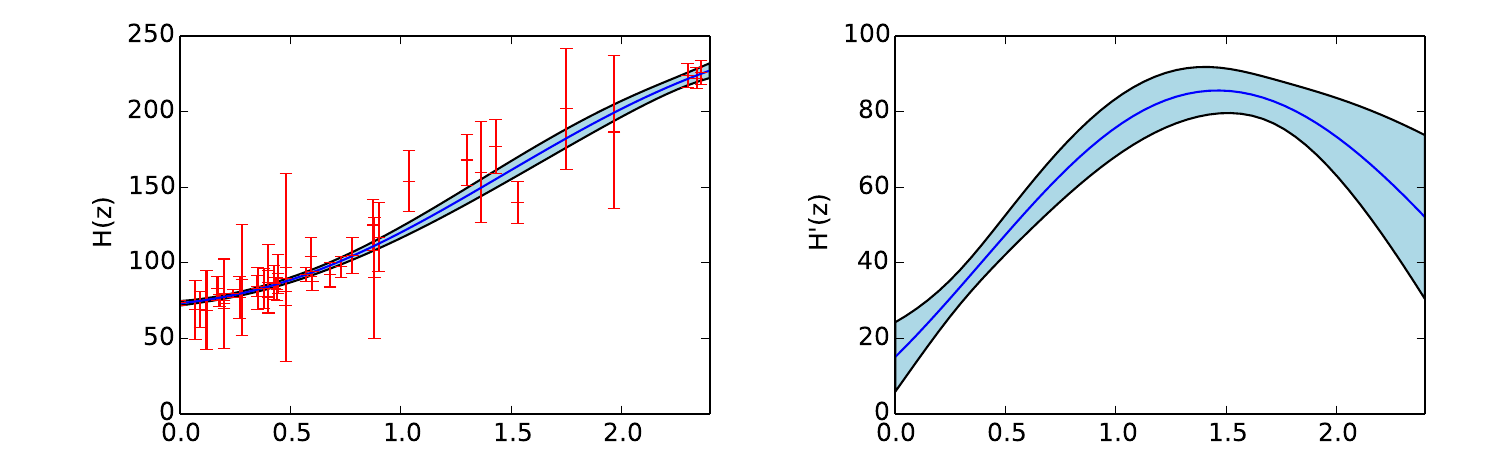} 
 \end{array}$
 \end{center}
\caption{GP reconstruction of $H(z)$ and $H^{\prime}(z)$ for the $40$-point sample given in Table \ref{tab:Table0}, when $H_{0} = 73.52 \pm 1.62$ reported by the Hubble mission. The top panel depicts the reconstruction when the kernel is that of Eq. (\ref{eq:kernel1}). The middle panel corresponds to the reconstruction from the kernel given by Eq. (\ref{eq:kernel3}). And the bottom panel is the results for the kernel given by Eq. (\ref{eq:kernel2}). The $^{\prime}$ means derivative with respect to the redshift $z$. }
 \label{fig:Fig0_0}
\end{figure}

The reconstruction of $H(z)$ and $H^{\prime}(z)$ when $H_{0} = 73.52 \pm 1.62$ km/s/Mpc is to be found in Fig. (\ref{fig:Fig0_0}), in the three cases when the kernel is given by Eq. (\ref{eq:kernel1}) (top panel), by Eq. (\ref{eq:kernel3}) (middle panel) and by Eq. (\ref{eq:kernel2}) (bottom panel), respectively.

\subsection{Model with $\omega_{dm} = \omega_{0}$}

The first model we have analyzed using a GP is the one where the dark matter dynamics is given by the following equation
\begin{equation}\label{eq:NCDM_1}
\rho_{dm} =   3 H^{2}_{0} \Omega^{(0)}_{dm} (1 + z)^{3 (1 + \omega_{0} )},
\end{equation}
where $H_{0}$ and $\Omega^{(0)}_{dm}$ are the Hubble parameter and the fraction of the dark matter at $z=0$, respectively. This is the model obtained from Eq. (\ref{eq:dm}) with $\omega_{0} \neq 0$ and $P_{dm} = \omega_{0} \rho_{dm}$, respectivley. Therefore, for $\rho_{de}$ we will have (see  Eq. (\ref{eq:F1}))
\begin{equation}\label{eq:NCDM_1_rhode}
\rho_{de} = 3 H^{2} - \rho_{dm} = 3H^{2} - 3 H^{2}_{0} \Omega^{(0)}_{dm} (1 + z)^{3 (1 + \omega_{0} )},
\end{equation}
from where it follows that $\rho^{\prime}_{de} = d \rho_{de}/dz$ is given by
\begin{equation}
\rho^{\prime}_{de} = 6 H H^{\prime} -9 H^2_{0} (1 + \omega_{0}) \Omega^{(0)}_{dm} (1 + z)^{3 \omega_{0}+2}.
\end{equation}
After some algebra, for $P_{de}$, Eq. (\ref{eq:Pde}), we  get
\begin{equation}
P_{de} = -3 H^2 - 3 H^{2}_{0} \omega_{0} \Omega^{(0)}_{dm} (1+ z)^{3 (1 + \omega_{0} )} + 2 (1 + z) H H^{\prime}.
\end{equation}
Having $P_{de}$ and $\rho_{de}$ expressed in terms of  $H$ and its derivatives, we see that it is possible to study the equation of state parameter for dark energy and reveal its nature when one has a deviation from the cold dark matter paradigm. In particular, for the deviation described by Eq. (\ref{eq:NCDM_1}), for $\omega_{de} = P_{de}/\rho_{de}$, one gets
\begin{equation}\label{eq:NCDM_1_omegade}
\omega_{de} = - \frac{3 H^2 + 3 H{^2}_{0} \omega_{0} \Omega^{(0)}_{dm} (1 + z)^{3 (1 + \omega_{0} )} - 2 (1 + z) H H^{\prime}} {3 H^2 - 3 H^{2}_{0} \Omega^{(0)}_{dm} (1 + z)^{3 (1 + \omega_{0}) }}.
\end{equation}
Actually, from the above equations, it follows that, in this case, only the Hubble function $H$ and its first-order derivative, $H^{\prime}$, are required to be reconstructed. The constraints on the parameters obtained for this case can be found in Table \ref{tab:Table1}. We see that the resulting constraints do hint towards a deviation from the cold dark matter paradigm encoded in the expansion rate data. In particular, for the $H_{0}$ value obtained for the reconstruction using only the available $H(z)$ data given in Table \ref{tab:Table0}, we have already found a hint that there is a deviation from the cold dark matter model (upper part of Table \ref{tab:Table1}). 

\begin{table}[ht]
	\centering
	
	\begin{tabular}{|c|c|c|c|c|} 
		\hline
		$Kernel$  & $\Omega^{(0)}_{dm}$ & $H_{0}$ & $\omega_{0}$\\
		\hline
		 Squared Exponent & $0.262 \pm 0.011$ & $71.286 \pm 3.743$ km/s/Mpc & $-0.069 \pm 0.011$  \\
		\hline
		Cauchy & $0.263 \pm 0.011$ & $71.472 \pm 3.879$ km/s/Mpc & $-0.075\pm 0.011$ \\
		\hline
		Matern $(\nu = 9/2)$ & $0.266 \pm 0.011$ & $71.119 \pm 3.867$ km/s/Mpc & $-0.075\pm 0.012$ \\
		\hline
     
     \multicolumn{3}{c}{} \\ \hline
     
      		Squared Exponent & $0.273 \pm 0.011$ & $73.52 \pm 1.62$ km/s/Mpc & $-0.075 \pm 0.012$  \\
		\hline
		Cauchy & $0.278 \pm 0.011$ & $73.52 \pm 1.62$ km/s/Mpc & $-0.083\pm 0.012$ \\
		\hline
		Matern $(\nu = 9/2)$ & $0.281 \pm 0.012$ & $73.52 \pm 1.62$ km/s/Mpc & $-0.088\pm 0.013$ \\
		\hline

	\multicolumn{3}{c}{} \\ \hline
	
		Squared Exponent & $0.293 \pm 0.011$ & $67.66 \pm 0.42$ km/s/Mpc & $-0.051 \pm 0.017$  \\
		\hline
		Cauchy & $0.285 \pm 0.015$ & $67.66 \pm 0.42$ km/s/Mpc & $-0.043\pm 0.015$ \\
		\hline
		Matern $(\nu = 9/2)$ & $0.291 \pm 0.013$ & $67.66 \pm 0.42$ km/s/Mpc & $-0.049\pm 0.017$ \\
		\hline

	\end{tabular}
	\caption{Constraints on the parameters for the cosmological model where the deviation from the cold dark matter case is given by Eq.~(\ref{eq:NCDM_1}) and $\omega_{0} \neq 0$. The constraints have been obtained for three kernels, Eqs. (\ref{eq:kernel1}),  (\ref{eq:kernel3}), and (\ref{eq:kernel2}), respectively. The upper part of the table stands for the case when the $H_{0}$ value has been predicted from the GP. In this case, we have found $\omega_{de} \in (-1.35, -0.69)$, with  mean $\omega_{de} \approx -1.06$, when the kernel is given by Eq. (\ref{eq:kernel1}). When the kernel corresponds to Eq. (\ref{eq:kernel3}), then $\omega_{de} \in (-1.37, -0.63)$ with  mean $\omega_{de} \approx -1.05$. Finally $\omega_{de} \in (-1.38, -0.65)$ with mean  $\omega_{de} \approx -1.06$ correspond to the kernel given by Eq. (\ref{eq:kernel2}), respectively. The middle part of the table stands for the case when the value $H_{0} = 73.52 \pm 1.62$ km/s/Mpc from the Hubble Space Telescope has been merged together with the available expansion rate data given in Table \ref{tab:Table0}, to reconstruct the values of $H(z)$ and $H^{\prime}(z)$. Finally, the lower part of the table stands for the case when the $H_{0} = 67.4 \pm 0.5$ km/s/Mpc from the Planck CMB data analysis has been merged with the available expansion rate data given in Table \ref{tab:Table0} and used in the reconstruction.}
	\label{tab:Table1}
\end{table} 

\begin{figure}[h!]
 \begin{center}$
 \begin{array}{cccc}
\includegraphics[width=80 mm]{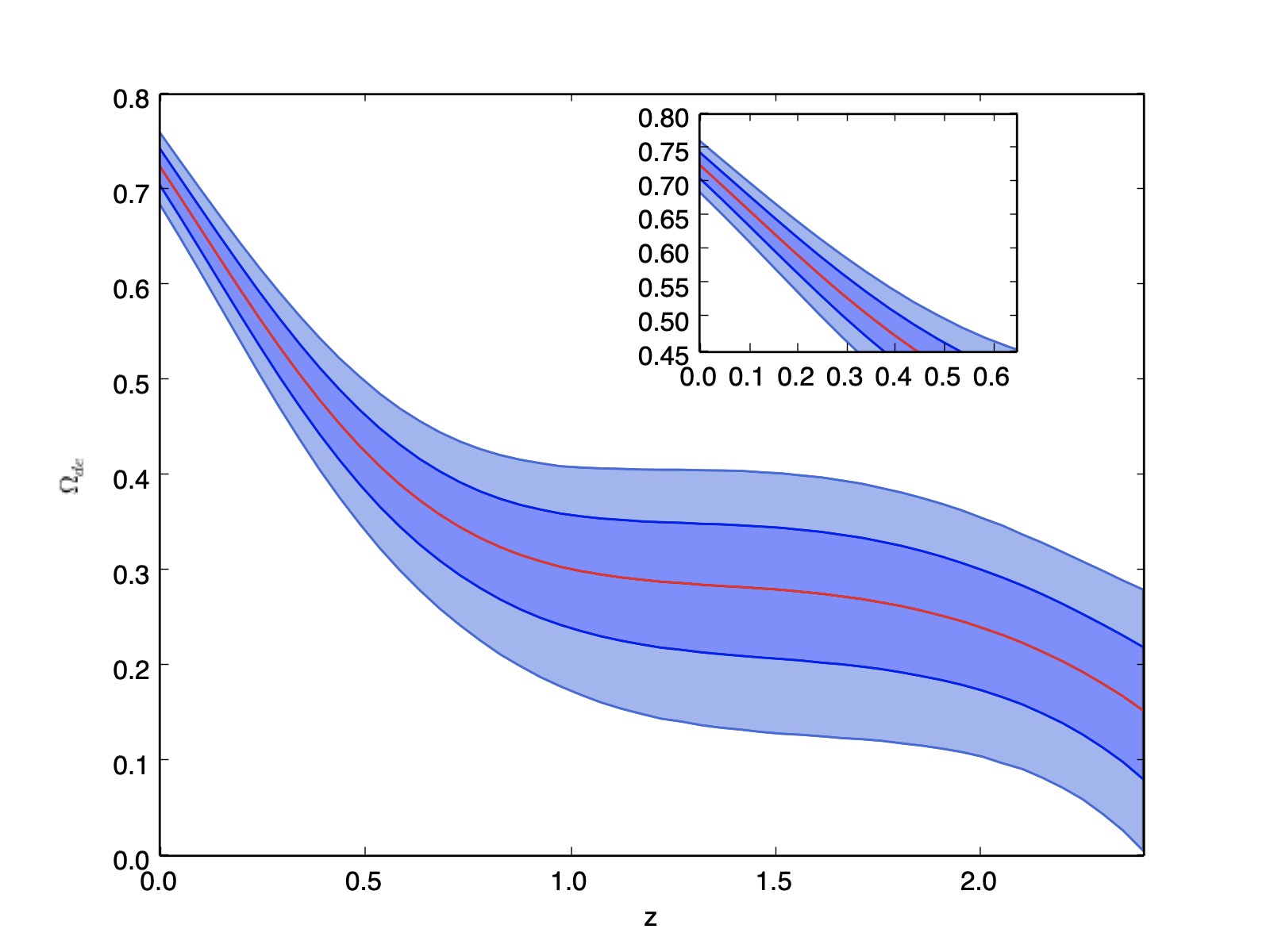} &
\includegraphics[width=80 mm]{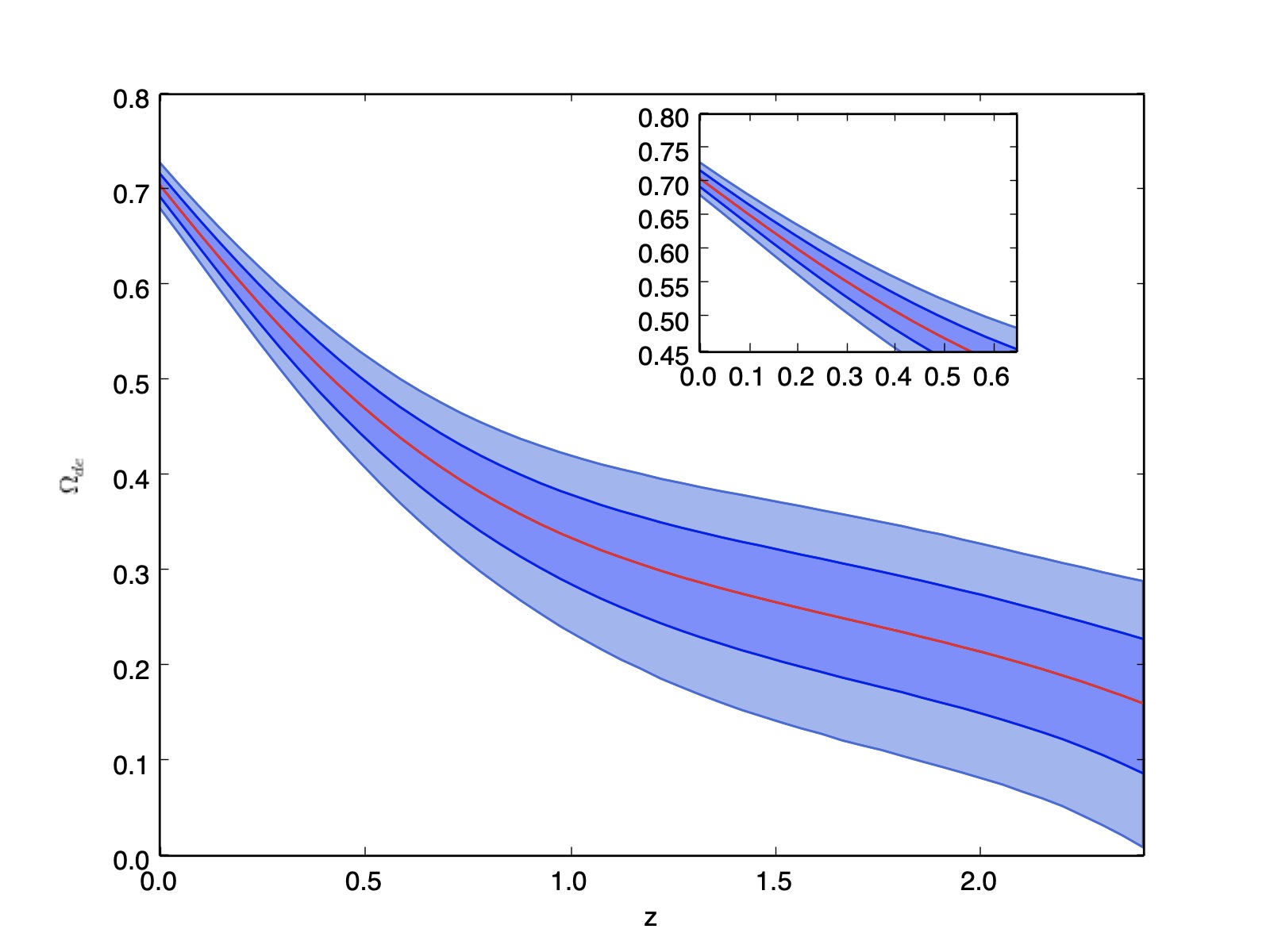} 
 \end{array}$
 \end{center}
\caption{GP reconstruction of $\Omega_{de} = \rho_{de}/3H^{2}$ for the model given by Eq.~(\ref{eq:NCDM_1}) and $\omega_{0} \neq 0$. The plot on the left-hand side corresponds to the reconstruction when $H_{0} =73.52 \pm 1.62$ km s$^{-1}$ Mpc$^{-1}$ has been merged to the expansion rate data used in the GP and the kernel is given by Eq. (\ref{eq:kernel1}). The right-hand side plot corresponds to the GP reconstruction when $H_{0} =67.40 \pm 0.5$ km s$^{-1}$ Mpc$^{-1}$ has been merged to the expansion rate data used in the GP and the kernel is given by Eq. (\ref{eq:kernel1}).}
 \label{fig:Fig0_1}
\end{figure}

\begin{figure}[h!]
 \begin{center}$
 \begin{array}{cccc}
\includegraphics[width=80 mm]{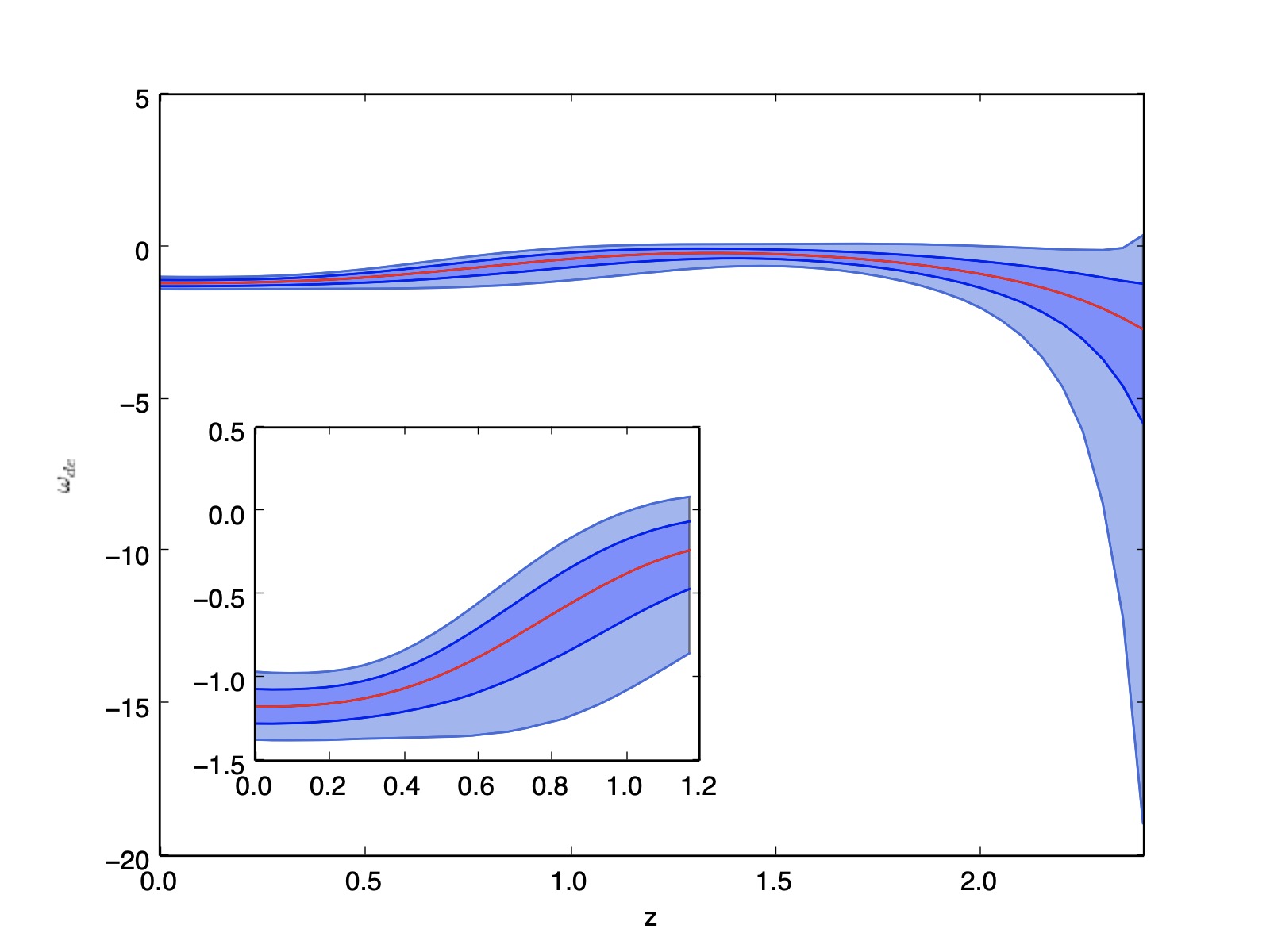} &
\includegraphics[width=80 mm]{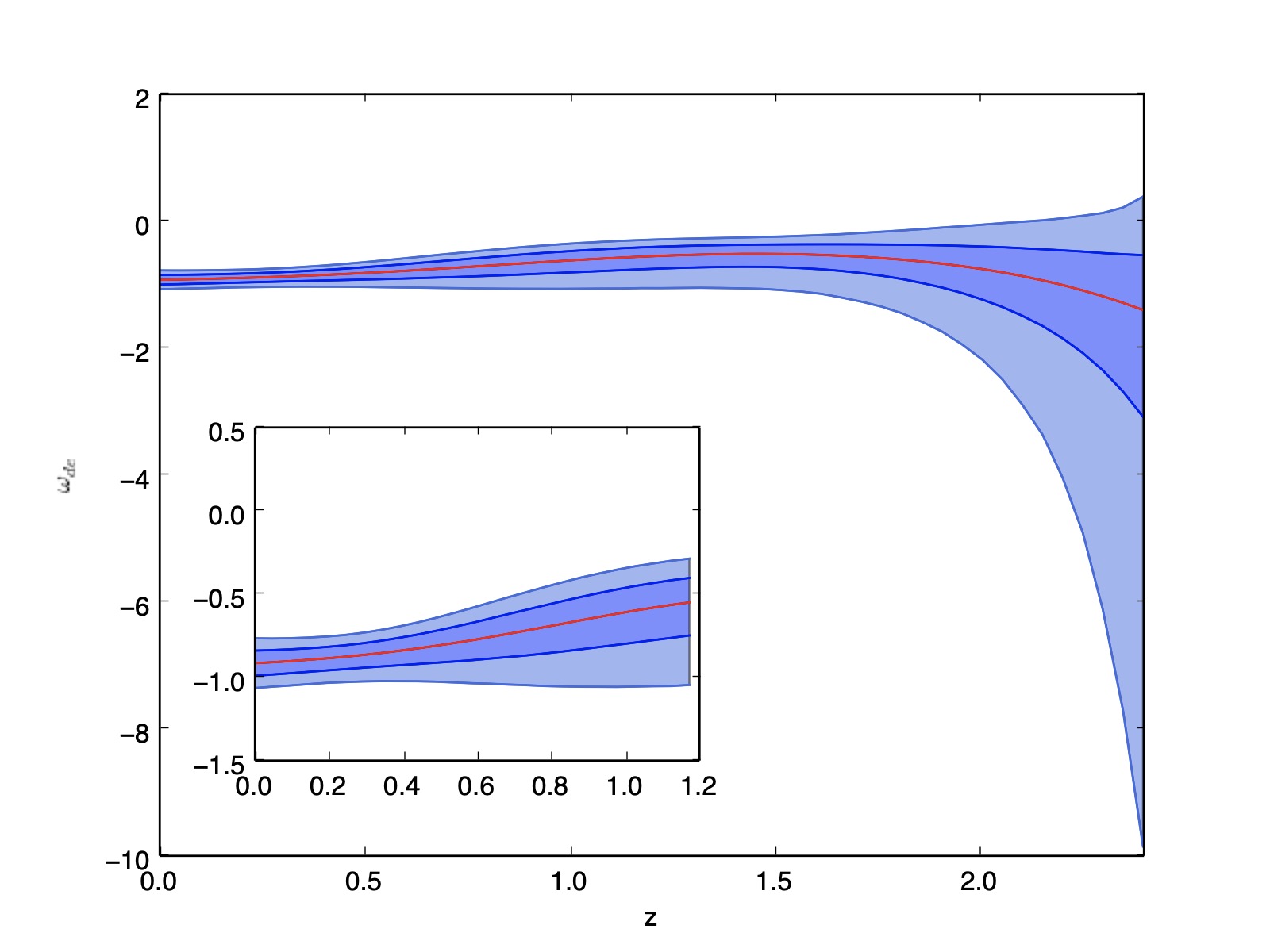} 
 \end{array}$
 \end{center}
\caption{GP reconstruction of $\omega_{de}$, Eq. (\ref{eq:NCDM_1_omegade}) for the model  given by Eq.~(\ref{eq:NCDM_1}) and $\omega_{0} \neq 0$. The plot on the left-hand side corresponds to the reconstruction when $H_{0} =73.52 \pm 1.62$ km s$^{-1}$ Mpc$^{-1}$ has been merged to the expansion rate data used in the GP and the kernel is given by Eq. (\ref{eq:kernel1}). In this case, $\omega_{de} \in (-1.37, -0.96)$ with mean $\omega_{de} \approx -1.17$ has been found. The right-hand side plot corresponds to the GP reconstruction when $H_{0} =67.40 \pm 0.5$ km s$^{-1}$ Mpc$^{-1}$ has been merged to the expansion rate data used in the GP and the kernel is given by Eq. (\ref{eq:kernel1}). In this case, $\omega_{de} \in (-1.06, -0.76)$ with mean $\omega_{de} \approx -0.91$  has been found. }
 \label{fig:Fig0_2}
\end{figure}

For the reconstruction of $H(z)$ and $H^{\prime}(z)$ we have used the value $H_{0} = 73.52 \pm 1.62$ km/s/Mpc coming from the Hubble Space Telescope with the available $H(z)$ data and again found a hint that there is a deviation from the cold dark matter model (middle part of Table \ref{tab:Table1}). Eventually, when we used the $H_{0} = 67.4 \pm 0.5$ km/s/Mpc (from the Planck CMB data analysis), we again found a similar hint as in the previous two cases (lower part of Table \ref{tab:Table1}). The strategy used here has allowed us to explore and estimate if and how the GP sees a connection between the $H_{0}$ tension problem and the deviation from the cold dark matter paradigm. In particular, a simple comparison of the constraints we have obtained shows that with $H_{0} = 67.4 \pm 0.5$ km/s/Mpc the deviation from the cold dark matter model should be smaller than when $H_{0} = 73.52 \pm 1.62$ km/s/Mpc. In other words, to solve the $H_{0}$ tension problem we need to have a strong deviation from the cold dark matter model according to the expansion rate data. We also noticed that the choice of the kernel can (strongly) affect the constraints on $\Omega^{(0)}_{dm}$ and $\omega_{0}$. However, this does not affect the main conclusion, namely that we find a deviation from the cold dark matter paradigm on cosmological scales. The results presented in Table \ref{tab:Table1}  hint towards possible new physics and deserves to be treated seriously. On the other hand, this confirms previously obtained results that were based on Bayesian Machine Learning processes \cite{Elizalde_H0}. 

The reconstruction of $\Omega_{de}$ and $\omega_{de}$ representing the behavior of the dark energy fraction and its equation of state when the kernel is given by Eq. (\ref{eq:kernel1}) can be found in Figs.~\ref{fig:Fig0_1} and \ref{fig:Fig0_2}, respectively. For brevity, we will discuss two cases, only, among all the ones we have studied, with similar qualitative results. They already shed light on the $H_{0}$ tension problem. 

In Fig.~\ref{fig:Fig0_1}, the left-hand side plot depicts the reconstructed behavior of the dark energy fraction $\Omega_{de}$ in a universe where the dark matter dynamics is given by Eq. (\ref{eq:NCDM_1}). In this case, $H_{0} = 73.52 \pm 1.62$ km/s/Mpc has been merged to the data given in Table \ref{tab:Table0} to be used in the reconstruction. There are various interesting scenarios that the model can reproduce. For instance, we see that in the evolutionary history of such universe, there is an epoch with $\Omega_{de} \approx 0$, which at $z=0$ yields a universe where $\Omega_{de} \approx 0.687$ (according to the lower bound of the reconstruction). On the other hand,  according to the upper bound of the reconstruction, there is also a possibility to obtain a model of the universe with an early dark energy component where the $H_{0}$ tension problem is solved. In this scenario, at $z = 0$ $\Omega_{de} \approx 0.762$. It should be mentioned that, in the recent literature, early dark energy models have been considered as an option to solve the $H_{0}$ tension problem. According to the mean of the reconstruction at $z=2.4$, we expect that $\Omega_{de} \approx 0.154$, while at $z = 0$, we have a model for the universe where $\Omega_{de} \approx 0.727$. Moreover, the reconstruction of $\Omega_{de}$ shows that various starting configurations and evolutionary paths leading to the recent dark energy-dominated universe can be realized within this model. Therefore, it is not surprising that a certain model-dependent parametrization (for instance, of the dark energy equation of state), will not be able to reveal a deviation from the cold dark matter model. Moreover, it is not excluded that similarly, the non-gravitational interaction for a given dark energy model can enter into the dynamics or be removed from the dynamics, respectively. The right hand side plot of Fig.~\ref{fig:Fig0_1} represents the $\Omega_{de}$ reconstruction when $H_{0} = 67.40 \pm 0.5$ km s$^{-1}$ Mpc$^{-1}$ has been used in the reconstruction process. It indicates that there is a difference in the dynamics of $\Omega_{de}$ as compared to the previous case, which strongly affects the constraints on $\Omega_{de}$ at $z=0$, too. It also affects the dynamics and the constraints on $\omega_{de}$ (Fig.~\ref{fig:Fig0_2}).

The analysis of $\omega_{de}$ shows that, with $H_{0} = 73.52 \pm 1.62$ km/s/Mpc, there is an epoch in the history of the universe where dark energy can be in a deep phantom phase, but during the evolution can change its nature becoming quintessence dark energy. On the other hand, it is not excluded that, starting from an $\omega_{de} > 0$, it could evolve and become quintessence dark energy (see the upper bound of the reconstruction on  Fig.~\ref{fig:Fig0_2}, left-hand side plot. We should stress that the analysis of $\omega_{de}$  hints that, within the considered scenario, a phantom crossing from above and from below can be realized. Moreover, the analysis of the dark energy equation of state does not reveal any strange or unusual behavior that would exclude a deviation from the cold dark matter model we have considered. During the analysis of this case, we found $\omega_{de} \in (-1.37, -0.96)$, with mean $\omega_{de} \approx -1.17$, for the kernel given by Eq. (\ref{eq:kernel1}). On the other hand, for the kernel of Eq. (\ref{eq:kernel3}), we obtained $\omega_{de} \in (-1.38, -0.94)$, with mean  $\omega_{de} \approx -1.16$, while $\omega_{de} \in (-1.38, -0.95)$, with mean $\omega_{de} \approx -1.17$ was found for the kernel given by Eq. (\ref{eq:kernel2}), respectively. 

Anyway, for $H_{0} = 67.40 \pm 0.5$ km s$^{-1}$ Mpc$^{-1}$ we got that in the recent universe, the quintessence nature of dark energy is preferable. On the other hand, the phantom crossing from above and from below is still possible. For this case,  $\omega_{de} \in (-1.06, -0.76)$ with  mean $\omega_{de} \approx -0.91$, for the kernel  given by Eq. (\ref{eq:kernel1}). Moreover, for the kernel of Eq. (\ref{eq:kernel3}), we found $\omega_{de} \in (-1.07, -0.74)$, with mean $\omega_{de} \approx -0.91$; while $\omega_{de} \in (-1.06, -0.74)$, with mean  $\omega_{de} \approx -0.91$, have been obtained for the kernel in Eq. (\ref{eq:kernel2}),  respectively. 

To end the discussion of this model, it should be mentioned that, in all the cases considered, the cosmological constant $\Lambda$ can be recovered.

\subsection{Model with $\omega_{m} = \omega_{0} + \omega_{1} z$}

This second model is considered to investigate the possibility that the deviation from the cold dark matter model has a dynamic nature. We start from the most simple case, namely the following linear model 
\begin{equation}\label{eq:omega_2}
\omega_{m} = \omega_{0} + \omega_{1} z
\end{equation}
The dynamics of this dark matter, given by $P_{dm} = (\omega_{0} + \omega_{1} z) \rho_{dm}$, are
\begin{equation}\label{eq:NCDM_2}
\rho_{dm} =   3 H^{2}_{0} \Omega^{(0)}_{dm} e^{3 \omega_{1} z} (1+ z)^{3 (1 + \omega_{0}-\omega_{1} )},
\end{equation}
with 
\begin{equation}\label{eq:drhodm_2}
\frac{d\rho_{dm}}{dz} = 9  H^{2}_{0} \Omega_{dm} e^{3 \omega_{1} z} (1 + z)^{3 \omega_{0}-3 \omega_{1}+2} (1 + \omega_{0}+\omega_{1} z)
\end{equation}
to be used to determine the dark energy pressure $P_{de}$. Moreover, after a simple algebra, for $\rho_{de}$ and $\rho^{\prime}_{de}$ we have
\begin{equation}\label{eq:rhode_1}
\rho_{de} = 3 H^{2} - \rho_{dm} = 3 H^{2} - 3 H^{2}_{0} \Omega^{(0)}_{dm} e^{3 \omega_{1} z} (1+ z)^{3 (1 + \omega_{0}-\omega_{1} )},
\end{equation}
and
\begin{equation}\label{eq:drhode_2}
\frac{d\rho_{de}}{dz} = 6 H H^{\prime} -9 H^{2}_{0} \Omega^{(0)}_{dm} e^{3 \omega_{1} z} (1 + z)^{3 \omega_{0}-3 \omega_{1}+2} (1 + \omega_{0}+\omega_{1} z),
\end{equation}
respectively. As a consequence, for the dark energy, 
\begin{equation}\label{eq:NCDM_2_omegade}
\omega_{de} = - \frac{3 H^{2}  + 3 H^{2}_{0} \Omega^{(0)}_{dm} e^{3 \omega_{1} z} (\omega_{0}+\omega_{1} z) (1 + z)^{3 (1 + \omega_{0}- \omega_{1} )} -  2 (1 + z) H H^{\prime} }{3 H^{2} - 3 H^{2}_{0} \Omega^{(0)}_{dm} e^{3 \omega_{1} z} (1 + z)^{3 (1 + \omega_{0} - \omega_{1} )}},
\end{equation}
since, from Eq. (\ref{eq:Pde}), for the pressure $P_{de}$, we have got
\begin{equation}
P_{de} = -3 H^{2} - 3 H^{2}_{0} \Omega^{(0)}_{dm} e^{3 \omega_{1} z} (\omega_{0}+\omega_{1} z) (1 + z)^{3 (1 + \omega_{0} -\omega_{1})} + 2 (1 + z) H H^{\prime} .
\end{equation}

The constraints on $\Omega^{(0)}_{dm}$, $\omega_{0}$ and $\omega_{1}$ can be found in Table \ref{tab:Table2}, from where we already see a clear deviation from the cold dark matter model. For this case, we just need now to reconstruct the Hubble function $H$ and its first-order derivative, $H^{\prime}$. The constraints on the model parameters --when the $H_{0}$ value has been obtained during the reconstruction using only available $H(z)$ data given in Table \ref{tab:Table0}-- can be found in the upper part of Table \ref{tab:Table2}. The middle part of Table \ref{tab:Table2} depicts the constraints for the case when, during the reconstruction of $H(z)$ and $H^{\prime}(z)$, one merges the $H_{0} = 73.52 \pm 1.62$ km/s/Mpc from the Hubble Space Telescope with the available $H(z)$ data. Finally, the lower part of Table \ref{tab:Table2} corresponds to the constraints in the case when, during the reconstruction, one merges the $H_{0} = 67.4 \pm 0.5$ km/s/Mpc (from the Planck CMB data analysis) with the available $H(z)$ data. 

As a first conclusion, the results indicate a clear deviation from the cold dark matter model and show also that this deviation might have a dynamic nature. Similar to the previous case, we see that for $H_{0} = 67.4 \pm 0.5$ km/s/Mpc, the deviation should be smaller than when $H_{0} = 73.52 \pm 1.62$ km/s/Mpc. However, the scenarios considered, including kernels, do not affect strongly the constraints on $\Omega^{(0)}_{dm}$, as it has been seen in the case of the first model.

\begin{table}[ht]
	\centering
	
	\begin{tabular}{|c|c|c|c|c|} 
		\hline
		$Kernel$  & $\Omega^{(0)}_{dm}$ & $H_{0}$ & $\omega_{0}$ & $\omega_{1}$\\
		\hline
		 Squared Exponent & $0.267 \pm 0.011$ & $71.286 \pm 3.743$ km/s/Mpc & $-0.074 \pm 0.011$  & $-0.015 \pm 0.005$ \\
		\hline
		Cauchy & $0.263 \pm 0.011$ & $71.472 \pm 3.879$ km/s/Mpc & $-0.074\pm 0.011$ & $-0.015 \pm 0.011$ \\
		\hline
		Matern $(\nu = 9/2)$ & $0.267 \pm 0.011$ & $71.119 \pm 3.867$ km/s/Mpc & $-0.074\pm 0.011$ & $-0.027 \pm 0.015$ \\
		\hline
     
     \multicolumn{4}{c}{} \\ \hline
     
      		Squared Exponent & $0.272 \pm 0.012$ & $73.52 \pm 1.62$ km/s/Mpc & $-0.065 \pm 0.012$  & $-0.014 \pm 0.011$ \\
		\hline
		Cauchy & $0.271 \pm 0.011$ & $73.52 \pm 1.62$ km/s/Mpc & $-0.065\pm 0.011$ & $-0.009 \pm 0.005$ \\
		\hline
		Matern $(\nu = 9/2)$ & $0.271 \pm 0.012$ & $73.52 \pm 1.62$ km/s/Mpc & $-0.067\pm 0.013$ & $-0.009 \pm 0.007$ \\
		\hline

	\multicolumn{4}{c}{} \\ \hline
	
		Squared Exponent & $0.271 \pm 0.013$ & $67.66 \pm 0.42$ km/s/Mpc & $-0.022 \pm 0.014$  & $-0.009 \pm 0.007$ \\
		\hline
		Cauchy & $0.273 \pm 0.013$ & $67.66 \pm 0.42$ km/s/Mpc & $-0.027 \pm 0.017$ & $-0.008 \pm 0.007$ \\
		\hline
		Matern $(\nu = 9/2)$ & $0.273 \pm 0.012$ & $67.66 \pm 0.42$ km/s/Mpc & $-0.027 \pm 0.015$ & $-0.009 \pm 0.007$ \\
		\hline

	\end{tabular}
	\caption{Constraints on the parameters for the cosmological model where the deviation from the cold dark matter case is described by Eqs.~(\ref{eq:NCDM_2}) and (\ref{eq:omega_2}), respectively. The constraints have been obtained for three kernels, Eqs. (\ref{eq:kernel1}),  (\ref{eq:kernel3}), and (\ref{eq:kernel2}), respectively. The upper part of the table stands for the case when the $H_{0}$ value has been predicted from the GP. In this case, we have found $\omega_{de} \in (-1.36, -0.68)$, with mean $\omega_{de} \approx -1.07$, when the kernel is given by Eq. (\ref{eq:kernel1}). For the kernel given by Eq. (\ref{eq:kernel3}), we found $\omega_{de} \in (-1.37, -0.64)$, with  mean $\omega_{de} \approx -1.05$, while for the kernel given by Eq. (\ref{eq:kernel2}), one gets $\omega_{de} \in (-1.38, -0.65)$ with mean $\omega_{de} \approx -1.06$, respectively. The middle part of the table stands for the case when $H_{0} = 73.52 \pm 1.62$ km/s/Mpc coming from the Hubble Space Telescope has been merged together with the available expansion rate data given in Table \ref{tab:Table0}, to reconstruct $H(z)$ and $H^{\prime}(z)$. Finally, the lower part of the table stands for the case when $H_{0} = 67.4 \pm 0.5$ km/s/Mpc, from the Planck CMB data analysis, was merged with the available expansion rate data given in Table \ref{tab:Table0}, and then used in the reconstruction.}
	\label{tab:Table2}
\end{table}

\begin{figure}[h!]
 \begin{center}$
 \begin{array}{cccc}
\includegraphics[width=80 mm]{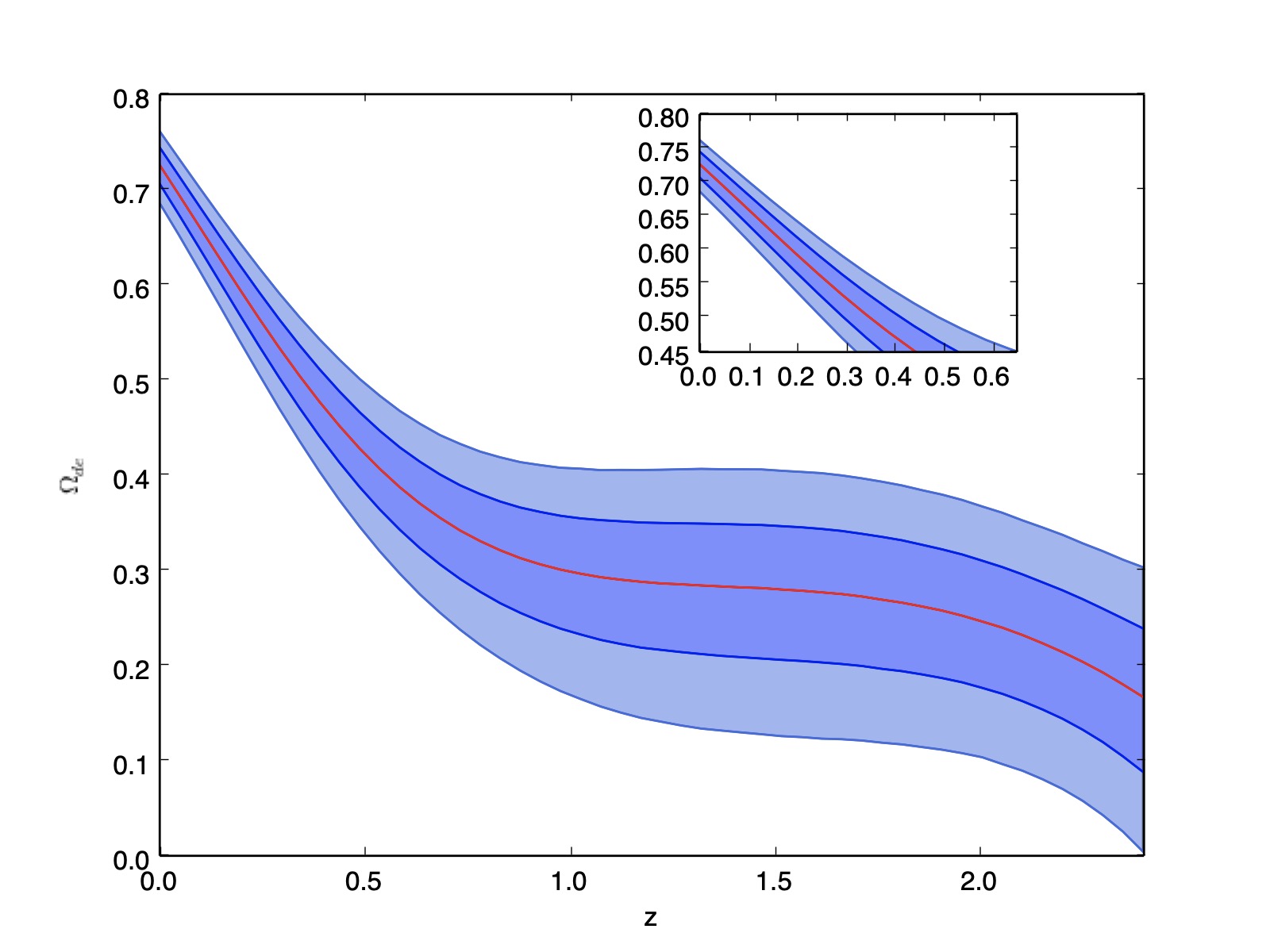} &
\includegraphics[width=80 mm]{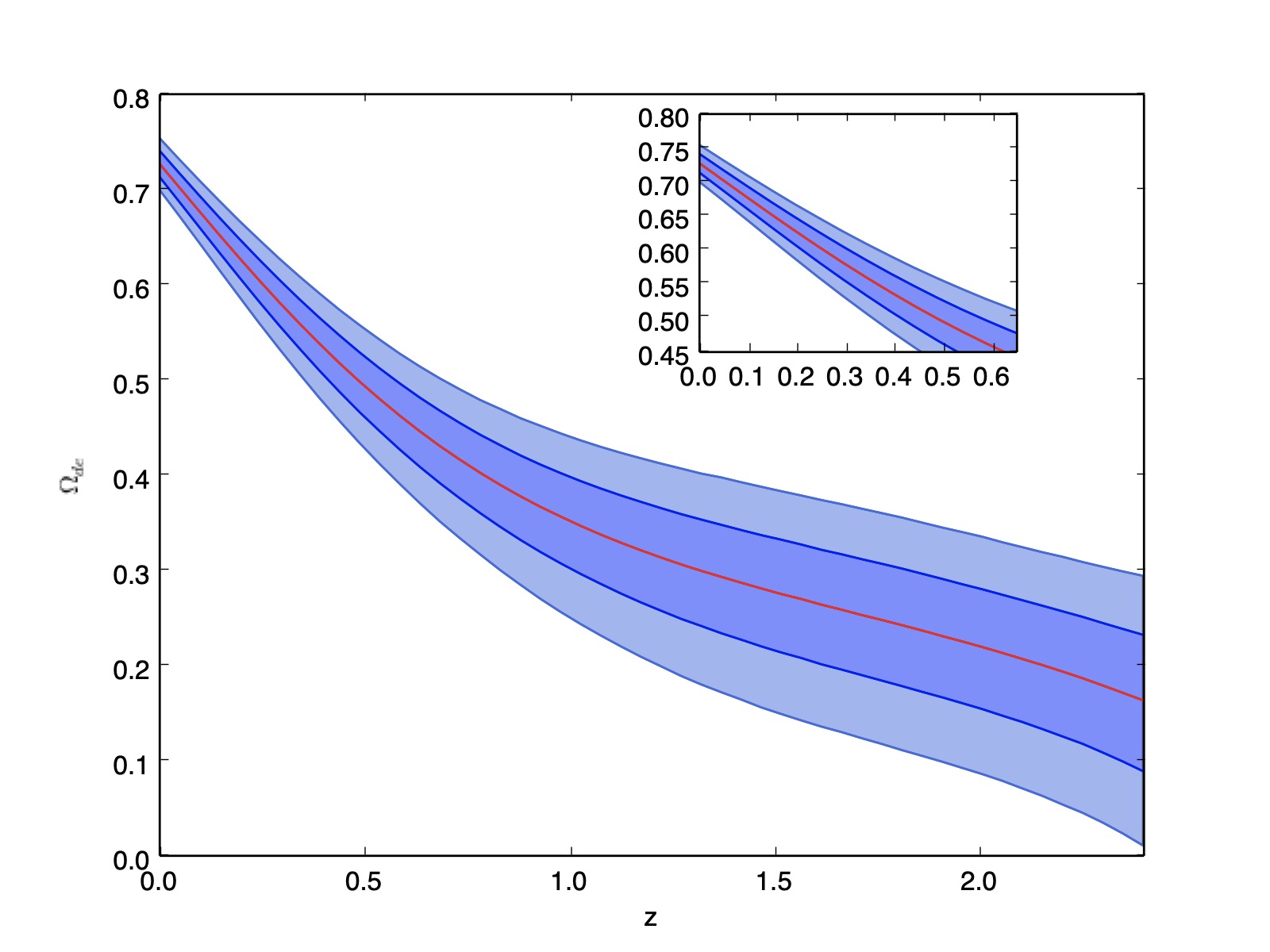} 
 \end{array}$
 \end{center}
\caption{GP reconstruction of $\Omega_{de} = \rho_{de}/3H^{2}$ for the model  given by Eqs. (\ref{eq:omega_2}) and (\ref{eq:NCDM_2}). The plot on the left-hand side corresponds to the reconstruction when $H_{0} =73.52 \pm 1.62$ km s$^{-1}$ Mpc$^{-1}$ has been merged with the expansion rate data used in the GP, and the kernel is given by Eq. (\ref{eq:kernel1}). The right-hand side plot corresponds to the GP reconstruction when $H_{0} = 67.40 \pm 0.5$ km s$^{-1}$ Mpc$^{-1}$ has been merged with the expansion rate data used in the GP and the kernel is given by Eq. (\ref{eq:kernel1}).}
 \label{fig:Fig1_1}
\end{figure}

\begin{figure}[h!]
 \begin{center}$
 \begin{array}{cccc}
\includegraphics[width=80 mm]{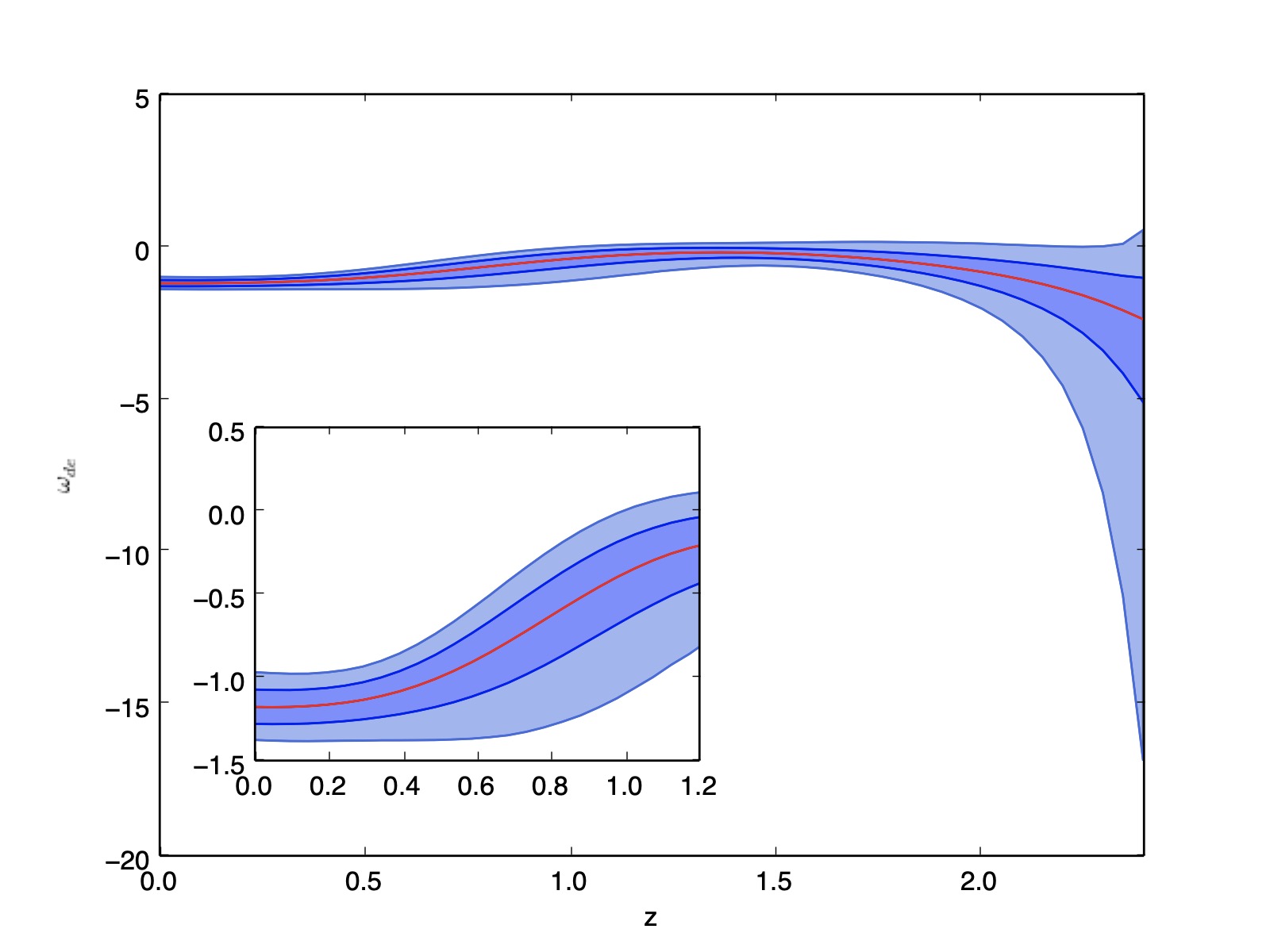} &
\includegraphics[width=80 mm]{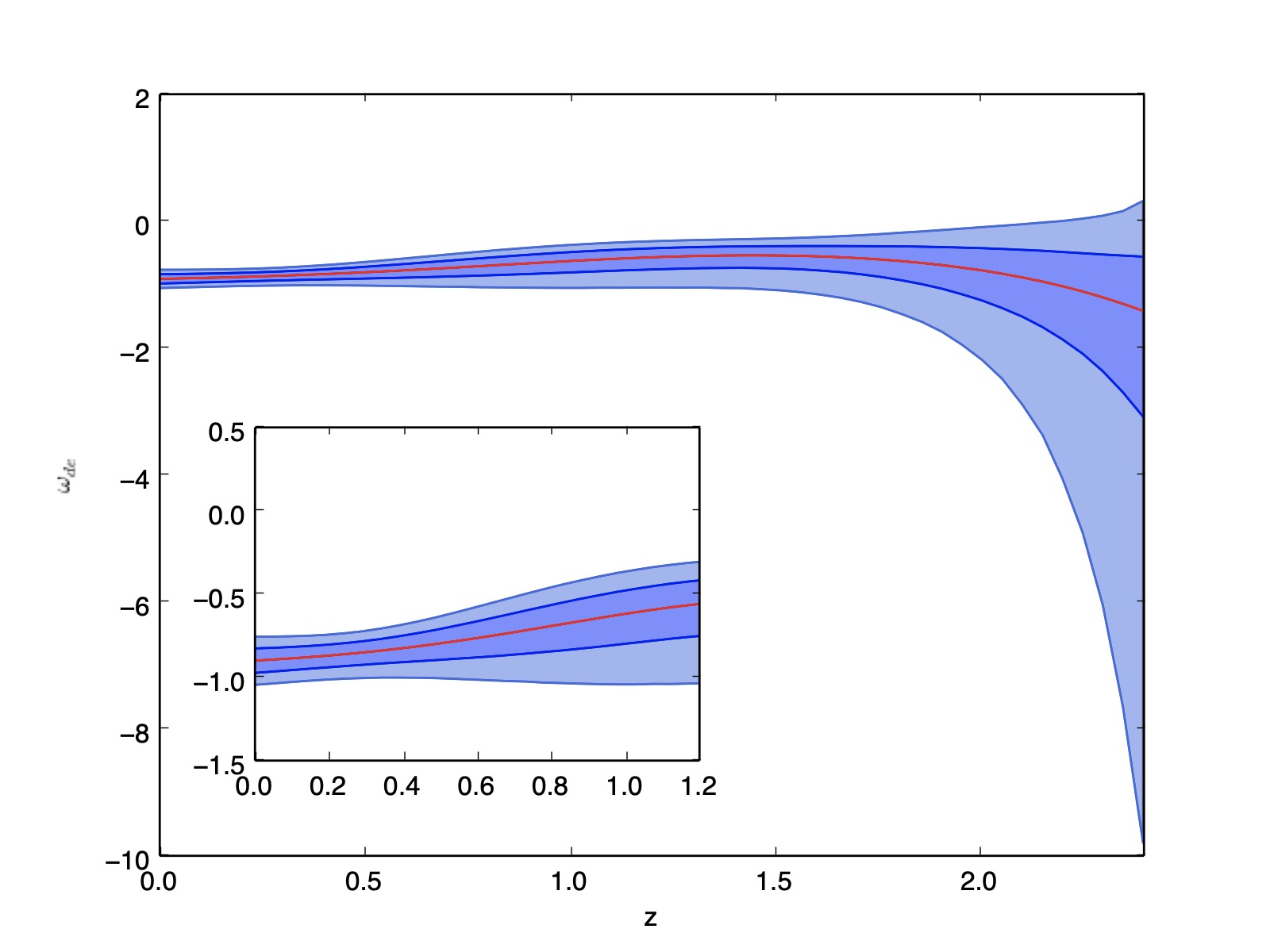} 
 \end{array}$
 \end{center}
\caption{GP reconstruction of $\omega_{de}$, Eq. (\ref{eq:NCDM_2_omegade}) for the model  given by Eqs. (\ref{eq:omega_2}) and (\ref{eq:NCDM_2}). The plot on the left-hand side corresponds to the reconstruction when $H_{0} =73.52 \pm 1.62$ km s$^{-1}$ Mpc$^{-1}$ has been merged with the expansion rate data used in the GP, and the kernel is given by Eq. (\ref{eq:kernel1}). In this case, one gets $\omega_{de} \in (-1.37, -0.96)$, with mean $\omega_{de} \approx -1.17$. The right-hand side plot corresponds to the GP reconstruction when $H_{0} = 67.40 \pm 0.5$ km s$^{-1}$ Mpc$^{-1}$ has been merged to the expansion rate data used in the GP and the kernel is given by Eq. (\ref{eq:kernel1}). In this case, the result is $\omega_{de} \in (-1.04, -0.75)$ with mean $\omega_{de} \approx -0.89$.}
 \label{fig:Fig1_2}
\end{figure}

The reconstruction of $\Omega_{de}$ and $\omega_{de}$, representing the behavior of the dark energy fraction and its equation of state, can be found in Fig.~\ref{fig:Fig1_1} and Fig.~\ref{fig:Fig1_2}, respectively, when the kernel used in the process is given by Eq. (\ref{eq:kernel1}). The left-hand side plot of Fig.~\ref{fig:Fig1_1} represents the reconstructed behavior of the dark energy fraction $\Omega_{de}$ in the universe where the dark matter dynamics is given by Eq. (\ref{eq:NCDM_2}) and $H_{0} = 73.52 \pm 1.62$ km/s/Mpc has been merged with the data given in Table \ref{tab:Table0}, to be used in the reconstruction. In this case, we have found $\omega_{de} \in (-1.37, -0.96)$ with mean $\omega_{de} \approx -1.17$ when the kernel is given by Eq. (\ref{eq:kernel1}). Correspondingly, for the kernel given by Eq. (\ref{eq:kernel3}), we found $\omega_{de} \in (-1.37, -0.94)$ with mean $\omega_{de} \approx -1.16$, while $\omega_{de} \in (-1.37, -0.94)$ with mean $\omega_{de} \approx -1.16$ has been found when the kernel is given by Eq. (\ref{eq:kernel2}), respectively. The right hand side plot of Fig.~\ref{fig:Fig1_1} corresponds to the case when $H_{0} =67.40 \pm 0.5$ km s$^{-1}$ Mpc$^{-1}$ has been used in the reconstruction process. It clearly indicates that, in both cases, there is a difference in the dynamics of $\Omega_{de}$, which strongly affects the constraints on $\Omega_{de}$ and $\omega_{de}$ at $z=0$. Again we need to stress that the analysis of the dark energy equation of state does not reveal any strange or unusual behavior that would exclude a deviation from the cold dark matter model. The fact that with $H_{0} = 67.40 \pm 0.5$ km s$^{-1}$ Mpc$^{-1}$ in the recent universe the quintessence nature of dark energy is preferable, can be used to dive deeper into the $H_{0}$ tension problem, which will be explored using other datasets in forthcoming papers (see the right-hand side plot of Fig.~\ref{fig:Fig1_2}). In this case, we found $\omega_{de} \in (-1.04, -0.75)$ with mean $\omega_{de} \approx -0.89$, when the kernel is given by Eq. (\ref{eq:kernel1}). When the kernel is that of Eq. (\ref{eq:kernel3}), we found $\omega_{de} \in (-1.05, -0.73)$ with mean $\omega_{de} \approx -0.89$. Finally, $\omega_{de} \in (-1.05, -0.73)$ with mean $\omega_{de} \approx -0.896$ has been found when the kernel is given by Eq. (\ref{eq:kernel2}). With the deviation from the cold dark matter model here considered, the cosmological constant $\Lambda$ can be recovered, yet. Indeed the reconstruction presented in Figs.~\ref{fig:Fig1_1} and \ref{fig:Fig1_2} shows that various scenarios can be reproduced, including early dark energy scenarios, and phantom crossing from above and below, respectively. The scenarios including the transition between different states of dark energy during the evolution process are also possible. Here, the cosmological constant $\Lambda$ can be recovered, too. 

\subsection{Model with $\omega_{m} = \omega_{0} + \omega_{1} z/(1+z)$ }

The last model we have analyzed should be seen as a further attempt to reveal the dynamic nature of the deviation, in this case through a non-linear model. The dark matter equation of state parameter has the following form 
\begin{equation}\label{eq:omega_3}
\omega_{m} = \omega_{0} + \frac{\omega_{1} z}{(1+z)},
\end{equation}
and the dark energy state equation reads 
\begin{equation}\label{eq:NCDM_3_omegade}
\omega_{de} = - \frac{ 3 H^{2} + 3 H_{0}^{2} \Omega^{(0)}_{dm} e^{-\frac{3 \omega_{1} z}{1 + z}} (\omega_{0}+ (\omega_{0} + \omega_{1}) z) (1 + z)^{3 \omega_{0}+3 \omega_{1}+2} -2 (z+1) H H^{\prime}} {3 H^{2} -3 H_{0}^{2} \Omega^{(0)}_{dm} e^{-\frac{3 \omega_{1} z}{1 + z}} (1+ z)^{3 (1 + \omega_{0}+\omega_{1})}},
\end{equation}
because, for the dark energy density dynamics according to Eq. (\ref{eq:Pde}), we already got
\begin{equation}\label{eq:NCDM_3}
\rho_{dm} = 3 H_{0}^{2} \Omega^{(0)}_{dm} e^{-\frac{3 \omega_{1} z}{1 + z}} (1 + z)^{3 (1 + \omega_{0}+ \omega_{1})},
\end{equation}
given that $P_{dm} = [ \omega_{0} + \omega_{1} z/(1+z)] \rho_{dm}$. The constraints on $\Omega^{(0)}_{dm}$, $\omega_{0}$ and $\omega_{1}$ can be found in Table \ref{tab:Table3}. In this case, too, we need only reconstruct the Hubble function $H$ and its first-order derivative, $H^{\prime}$. The constraints on the model parameters, when $H_{0}$ has been obtained during the reconstruction using only available $H(z)$ data given in Table \ref{tab:Table0}, can be found in the upper part of Table \ref{tab:Table3}. The middle part of Table \ref{tab:Table3} represents the constraints when, during the reconstruction of $H(z)$ and $H^{\prime}(z)$, we merged the $H_{0} = 73.52 \pm 1.62$ km/s/Mpc from the Hubble Space Telescope with the available $H(z)$ data. Finally, the lower part of Table \ref{tab:Table3} depicts the constraints when, during the reconstruction,  the $H_{0} = 67.4 \pm 0.5$ km/s/Mpc (from the Planck CMB data analysis) is merged with the available $H(z)$ data. In this case, too, the constraints we have obtained indicate a deviation from the cold dark matter model. And, similar to the previous cases, with $H_{0} = 67.4 \pm 0.5$ km/s/Mpc the deviation looks to be smaller than when $H_{0} = 73.52 \pm 1.62$ km/s/Mpc. 

\begin{table}[ht]
	\centering
	
	\begin{tabular}{|c|c|c|c|c|} 
		\hline
		$Kernel$  & $\Omega^{(0)}_{dm}$ & $H_{0}$ & $\omega_{0}$ & $\omega_{1}$\\
		\hline
		 Squared Exponent & $0.266 \pm 0.011$ & $71.286 \pm 3.743$ km/s/Mpc & $-0.068 \pm 0.011$  & $-0.024 \pm 0.011$ \\
		\hline
		Cauchy & $0.267 \pm 0.011$ & $71.472 \pm 3.879$ km/s/Mpc & $-0.067\pm 0.012$ & $-0.025 \pm 0.011$ \\
		\hline
		Matern $(\nu = 9/2)$ & $0.266 \pm 0.011$ & $71.119 \pm 3.867$ km/s/Mpc & $-0.068\pm 0.012$ & $-0.023 \pm 0.012$ \\
		\hline
     
     \multicolumn{4}{c}{} \\ \hline
     
      		Squared Exponent & $0.272 \pm 0.011$ & $73.52 \pm 1.62$ km/s/Mpc & $-0.064 \pm 0.011$  & $-0.022 \pm 0.011$ \\
		\hline
		Cauchy & $0.272 \pm 0.011$ & $73.52 \pm 1.62$ km/s/Mpc & $-0.069\pm 0.011$ & $-0.016 \pm 0.012$ \\
		\hline
		Matern $(\nu = 9/2)$ & $0.274 \pm 0.011$ & $73.52 \pm 1.62$ km/s/Mpc & $-0.069\pm 0.012$ & $-0.018 \pm 0.012$ \\
		\hline

	\multicolumn{4}{c}{} \\ \hline
	
		Squared Exponent & $0.274 \pm 0.012$ & $67.66 \pm 0.42$ km/s/Mpc & $-0.022 \pm 0.012$  & $-0.014 \pm 0.012$ \\
		\hline
		Cauchy & $0.273 \pm 0.011$ & $67.66 \pm 0.42$ km/s/Mpc & $-0.029 \pm 0.015$ & $-0.009 \pm 0.006$ \\
		\hline
		Matern $(\nu = 9/2)$ & $0.274 \pm 0.012$ & $67.66 \pm 0.42$ km/s/Mpc & $-0.022 \pm 0.012$ & $-0.014 \pm 0.012$ \\
		\hline

	\end{tabular}
	\caption{Constraints on the parameters for the cosmological model where the deviation from the cold dark matter is described by Eqs.~(\ref{eq:NCDM_3}) and (\ref{eq:omega_3}), respectively. The constraints have been obtained for three kernels, Eqs. (\ref{eq:kernel1}), (\ref{eq:kernel3}), and  (\ref{eq:kernel2}), and for three different values of the parameter $H_{0}$. The upper part of the table stands for the case when the value of $H_{0}$  has been predicted from a GP. In this case, we have found $\omega_{de} \in (-1.36, -0.69)$, with mean $\omega_{de} \approx -1.07$, for the case when the kernel is given by Eq. (\ref{eq:kernel1}). For the kernel given by Eq. (\ref{eq:kernel3}), we have got $\omega_{de} \in (-1.37, -0.64)$, with mean  $\omega_{de} \approx -1.05$, while $\omega_{de} \in (-1.39, -0.66)$ with mean $\omega_{de} \approx -1.07$ are the results when the kernel is given by Eq. (\ref{eq:kernel2}), respectively. The middle part of the table stands for the case when  $H_{0} = 73.52 \pm 1.62$ km/s/Mpc from the Hubble Space Telescope and the available expansion rate data given in Table \ref{tab:Table0}) have been merged, to reconstruct  $H(z)$ and $H^{\prime}(z)$. Finally, the lower part of the table stands for the case when  $H_{0} = 67.4 \pm 0.5$ km/s/Mpc, coming from the Planck CMB data analysis, has been used in the reconstruction.}
	\label{tab:Table3}
\end{table}

\begin{figure}[h!]
 \begin{center}$
 \begin{array}{cccc}
\includegraphics[width=80 mm]{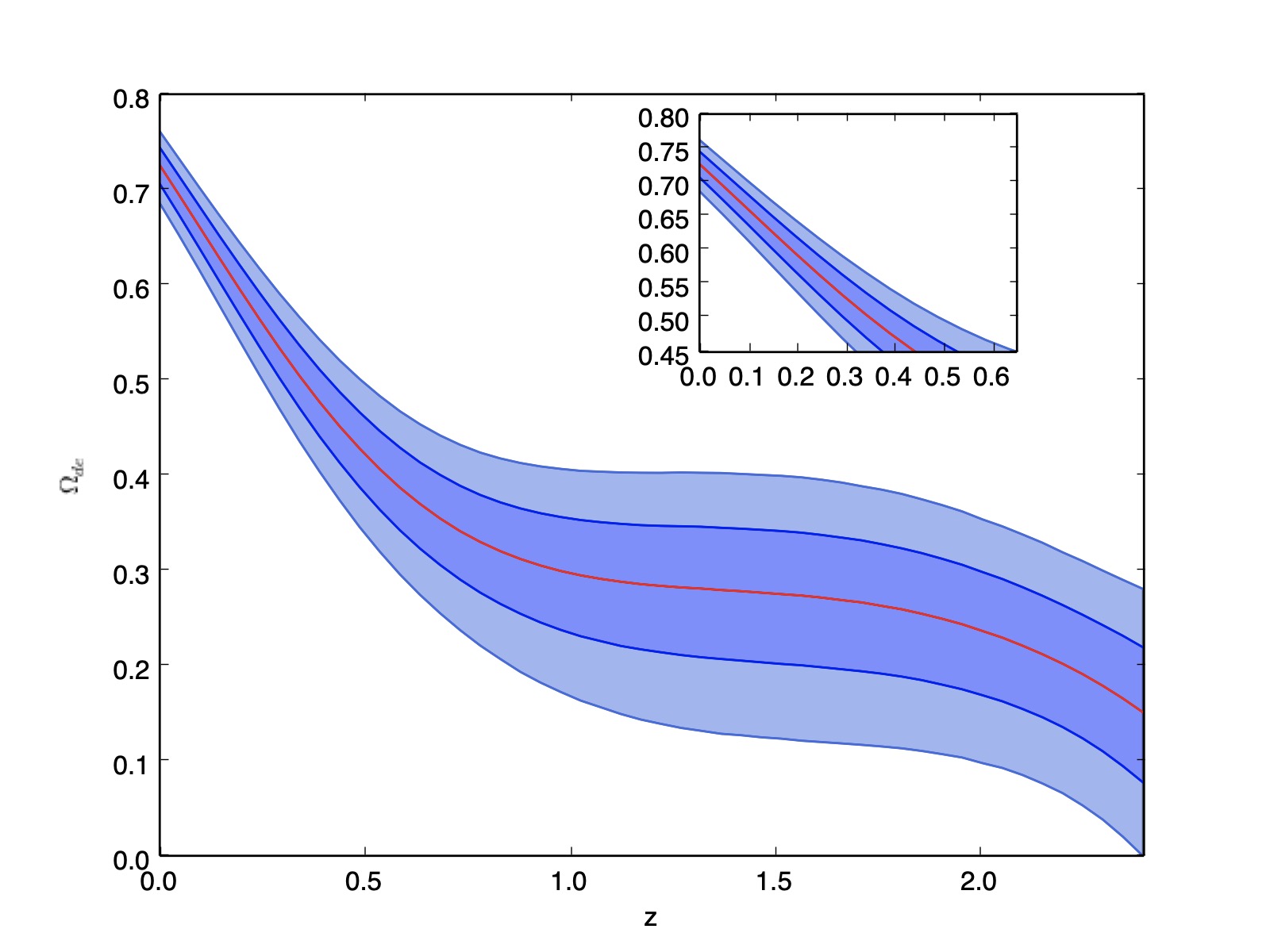} &
\includegraphics[width=80 mm]{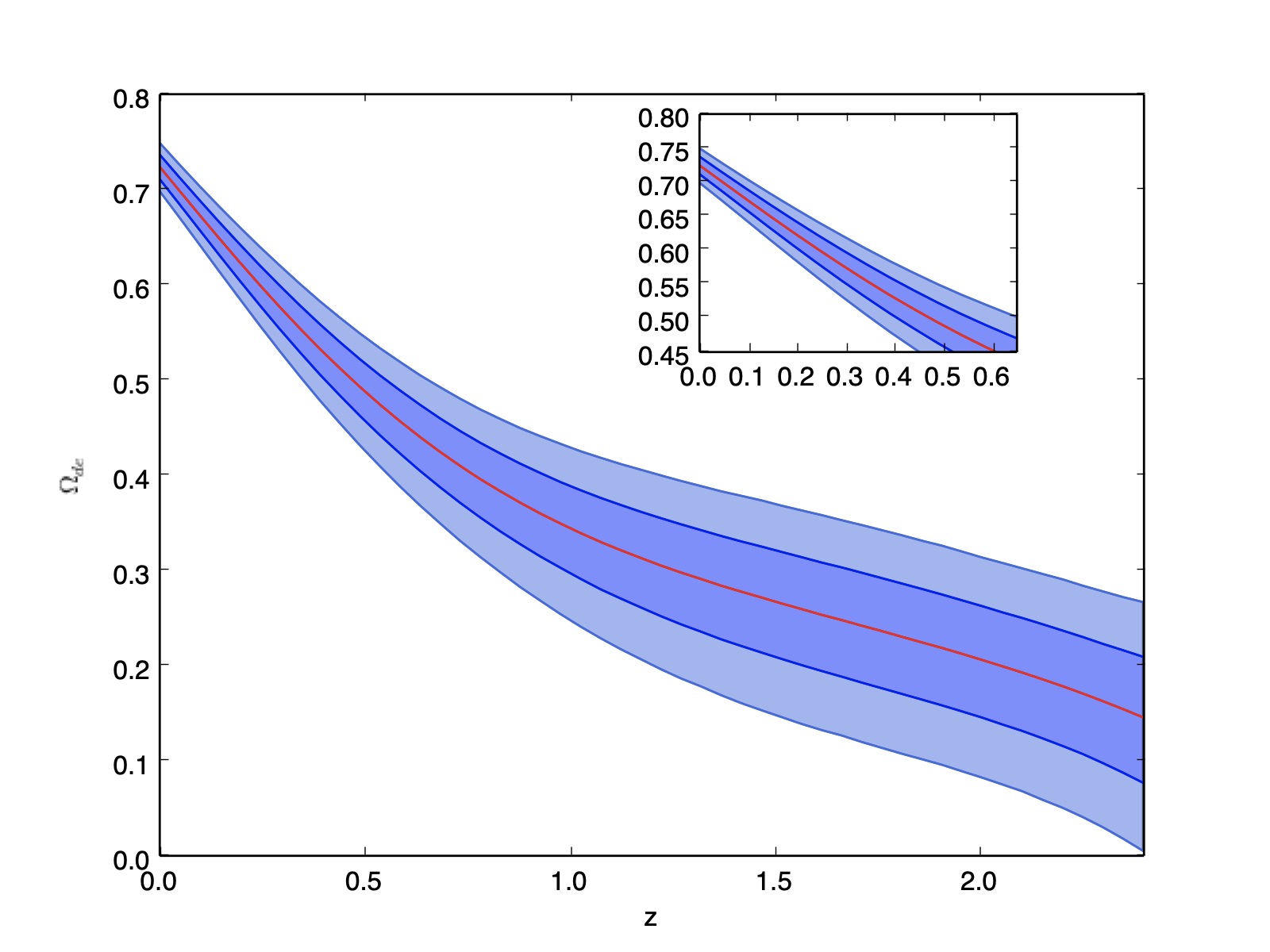} 
 \end{array}$
 \end{center}
\caption{GP reconstruction of $\Omega_{de} = \rho_{de}/3H^{2}$ for   $30$ samples of $H(z)$  deduced from the differential age method with $10$ additional samples obtained from the radial BAO method (Table \ref{tab:Table0}), when the model is given by Eqs. (\ref{eq:omega_3}) and (\ref{eq:NCDM_3}). The plot on the left-hand side represents the reconstruction when $H_{0} =73.52 \pm 1.62$ km s$^{-1}$ Mpc$^{-1}$ has been merged with the expansion rate data used in the GP, and the kernel is given by Eq. (\ref{eq:kernel1}). The right-hand side plot corresponds to the GP reconstruction when $H_{0} =67.40 \pm 0.5$ km s$^{-1}$ Mpc$^{-1}$ has been merged with the expansion rate data used in the GP, and the kernel is given by Eq. (\ref{eq:kernel1}).}
 \label{fig:Fig2_1}
\end{figure}

\begin{figure}[h!]
 \begin{center}$
 \begin{array}{cccc}
\includegraphics[width=80 mm]{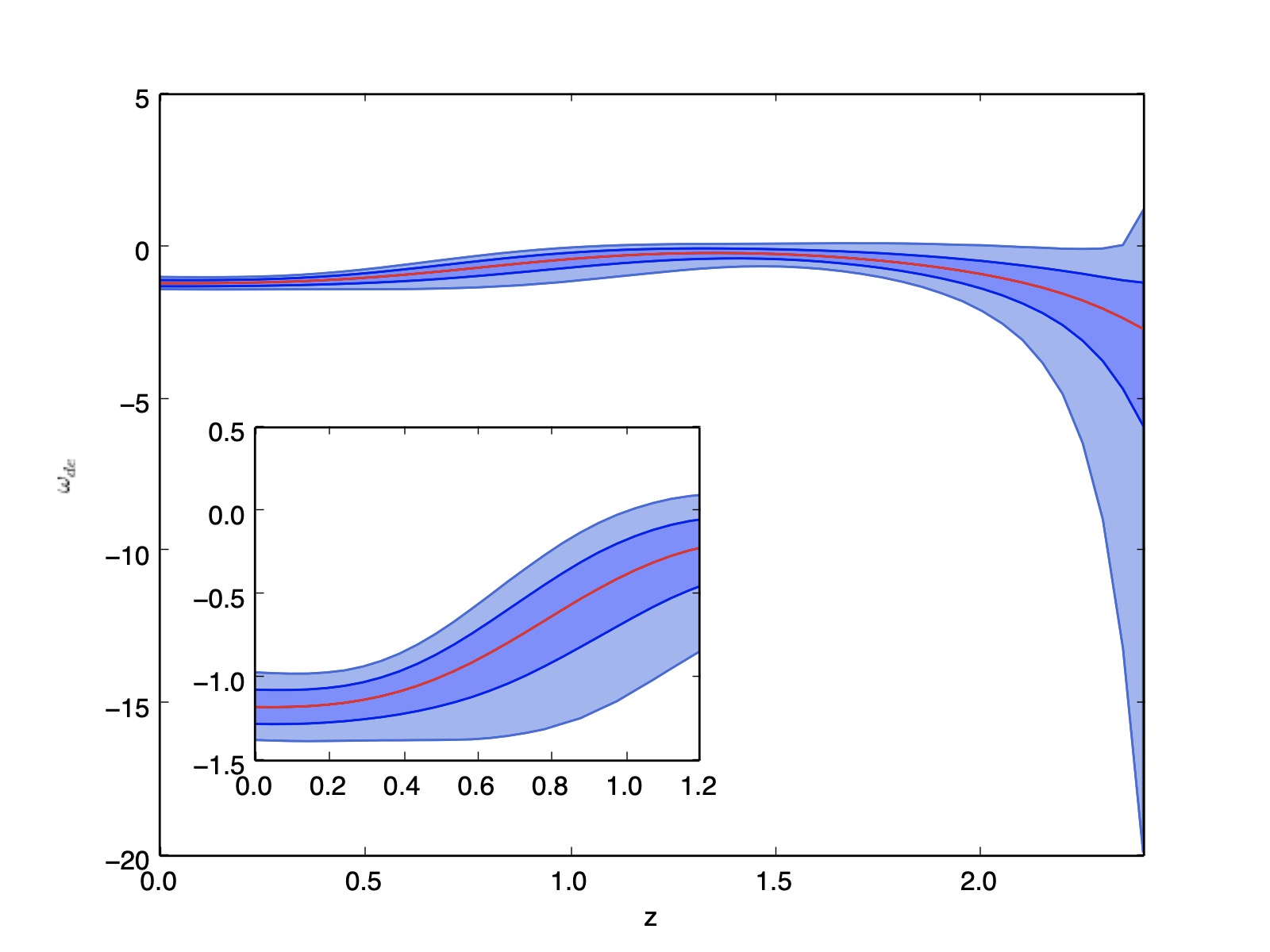} &
\includegraphics[width=80 mm]{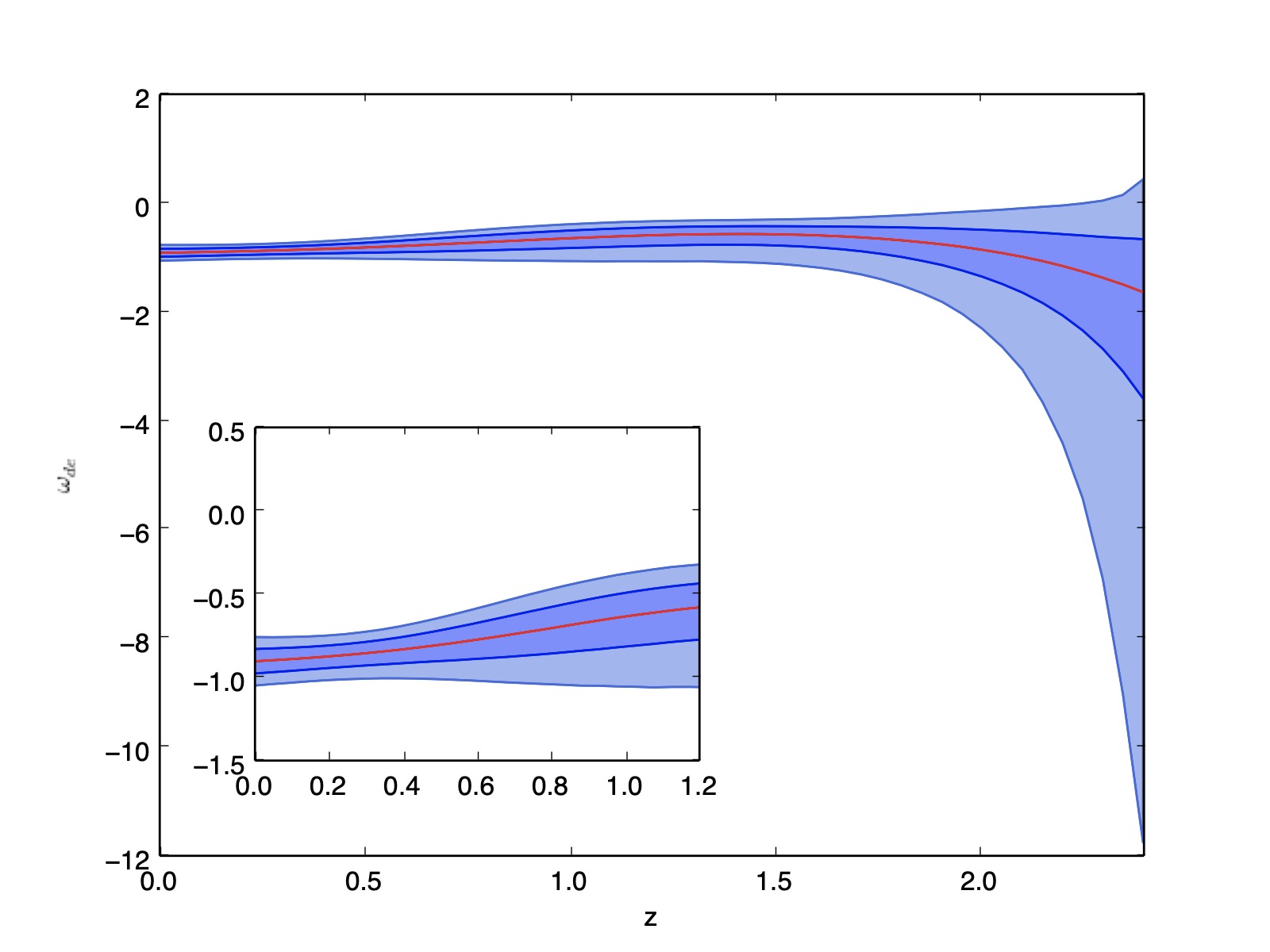} 
 \end{array}$
 \end{center}
\caption{GP reconstruction of $\omega_{de}$, Eq. (\ref{eq:NCDM_3_omegade}), for the $30$ $H(z)$ samples deduced from the differential age method with $10$ additional  samples obtained from the radial BAO method (Table \ref{tab:Table0}), when the model is given by Eqs. (\ref{eq:omega_3}) and (\ref{eq:NCDM_3}). The plot on the left-hand side represents the reconstruction when $H_{0} =73.52 \pm 1.62$ km s$^{-1}$ Mpc$^{-1}$ has been merged with the expansion rate data used in the GP, and the kernel is given by Eq. (\ref{eq:kernel1}). In this case, $\omega_{de} \in (-1.37, -0.96)$, with mean $\omega_{de} \approx -1.17$, has been found. The right-hand side plot corresponds to the GP reconstruction when $H_{0} = 67.40 \pm 0.5$ km s$^{-1}$ Mpc$^{-1}$ has been merged with the expansion rate data used in the GP and the kernel is given by Eq. (\ref{eq:kernel1}). In this case, we have obtained $\omega_{de} \in (-1.04, -0.75)$ with mean $\omega_{de} \approx -0.89$.}
 \label{fig:Fig2_2}
\end{figure}

The reconstruction of $\Omega_{de}$ and $\omega_{de}$, representing the behavior of the dark energy fraction and its equation of state for this model, are given in Figs.~\ref{fig:Fig2_1} and \ref{fig:Fig2_2}, respectively, when the kernel used in the process is the one of Eq. (\ref{eq:kernel1}). The left-hand side plot in Fig.~\ref{fig:Fig2_1} corresponds to the reconstructed behavior of the dark energy fraction $\Omega_{de}$ in the universe where the dark matter dynamics is that of Eq. (\ref{eq:NCDM_2}) and $H_{0} = 73.52 \pm 1.62$ km/s/Mpc has been merged with the data given in Table \ref{tab:Table0} to be used in the reconstruction. In this case, we have found $\omega_{de} \in (-1.37, -0.96)$ with  mean $\omega_{de} \approx -1.17$, when the kernel is given by Eq. (\ref{eq:kernel1}). For the kernel in Eq. (\ref{eq:kernel3}), we got $\omega_{de} \in (-1.37, -0.94)$ with mean $\omega_{de} \approx -1.16$, and for the kernel of Eq. (\ref{eq:kernel2}), $\omega_{de} \in (-1.37, -0.94)$, with mean $\omega_{de} \approx -1.16$.

The right hand side plot of Fig.~\ref{fig:Fig2_1} is the result of the $\Omega_{de}$ reconstruction when the value $H_{0} = 67.40 \pm 0.5$ km s$^{-1}$ Mpc$^{-1}$ has been used in the reconstruction process. Among other results that will be shortly presented below, we have found that $\omega_{de} \in (-1.04, -0.75)$, with mean $\omega_{de} \approx -0.89$, for the kernel given by Eq. (\ref{eq:kernel1});   $\omega_{de} \in (-1.05, -0.73)$ with mean $\omega_{de} \approx -0.89$, for the kernel of Eq. (\ref{eq:kernel3}); while $\omega_{de} \in (-1.05, -0.74)$, with mean $\omega_{de} \approx -0.89$ has been found for the kernel given by Eq. (\ref{eq:kernel2}), respectively. Results clearly indicate that, in both cases, there is a difference in the dynamics of $\Omega_{de}$, which strongly affects the constraints on $\Omega_{de}$ and $\omega_{de}$ at $z=0$, too. The analysis of $\omega_{de}$ shows that for $H_{0} = 73.52 \pm 1.62$ km/s/Mpc in the history of the universe, there is an epoch where dark energy can be in a deep phantom phase but during the evolution can change its nature becoming quintessence dark energy. The model here considered allows for the realization of different scenarios, similarly as in previous cases. The reconstructions obtained clearly point out qualitative similarities with the previous cases, which have been already discussed before. 

To finish this section, we need to stress that, choosing a preferable model among the ones considered, by just using a GP and expansion rate data,  is indeed a hard process. Additional statistical tools should be applied to perform the model selection with guarantees. More work in this promising direction is needed, as will be discussed in the next section  \ref{sec:Discussion}. 

\section{\large{Discussion}}\label{sec:Discussion}

In this paper, we have used GPs to reveal a possible connection between the $H_{0}$ tension problem and its solution in terms of a deviation from the cold dark matter model. This extends previous work, where hints about such a deviation was obtained using a Bayesian Machine Learning approach. There, the learning method was based on the generated expansion rate, which already indicated that a deviation from the cold dark matter paradigm might solve the $H_{0}$ problem. Moreover, a sound possibility that the mentioned deviation could replace a proposed non-gravitational interaction between dark energy and dark matter was also put forward (see \cite{Elizalde_H0}). In this regard, a mere deviation from the cold dark matter paradigm may be considered more natural, instead of having to rely on a mechanism that generates interaction between two energy sources, which have different acting scales and roles in our universe. 

Improving that work, we have here used a pure GP and real, available expansion rate data, at the expense of significantly restricting the redshift range of the analysis. Results in this case are far more reliable and we need to stress that, despite the restricted range, also here hints indicating a deviation from the cold dark matter model have been distinctly inferred. In our analysis, we have used $40$-point $H(z)$ data consisting of $30$-point samples deduced from the differential age method and $10$-point samples obtained from the radial BAO method. Already in this prospective situation, we have found a hint that the deviation from the cold dark matter paradigm could be a source for the $H_{0}$ tension problem. Moreover, we have learned that, when the deviation is described by a free parameter, $\omega_{0} \neq 0$, to solve the $H_{0}$ tension problem, we would need a quite strong deviation from the cold dark matter model. 

The consideration of the other two models with $\omega(z)_{dm}$ has confirmed this connection. Using three different kernels and three different values for the $H_{0}$ parameter, we have been able to reveal that an early dark energy component may solve the $H_{0}$ tension problem. In this case, the early dark energy model originates from the mentioned deviation, without the need to introduce auxiliary dark energy models or non-gravitational interactions of any sort. On the other hand, we have also learned, from the reconstruction, that phantom crossing from above (and/or from below) can be used to craft a solution for the $H_{0}$ tension problem, too. 

For $H_{0} = 67.40 \pm 0.5$ km s$^{-1}$ Mpc$^{-1}$ an indication that the quintessence nature of dark energy is preferable has been obtained. We should also stress the fact that, in all cases under study, the cosmological constant $\Lambda$ could be recovered, with the independence of the chosen scenario deployed in our study. It is not according to the mean of the reconstruction, therefore, we have a hint that the $H_{0}$ tension still indicates a problem with the physics, indeed.

Finally, we should stress, once more, that the behavior we have obtained for $\Omega_{de}$ (and $\Omega_{dm}$, too) and for $\omega_{de}$ does not contain any indication that the deviation from the cold dark matter model is an artifact, originating in the procedure employed itself. Rather, the results of the present study provide a clear image of the fact that the deviation is indeed imprinted into the observational expansion rate data. What is more, we prove that the deviation can most likely have a dynamical component, and two possibilities thereof have been explored. 

The results here obtained, combined with the ones reported in \cite{Elizalde_H0},  provide support for an interesting, alternative way to approach the solution to various fundamental problems of modern cosmology. They already support the idea of extending the analysis in various directions. The most important issue would be to understand how the deviation from the cold dark matter model affects structure formation in our universe. It is key to understand, in particular, how it may affect the constraints on neutrino physics and help (or prevent) a unified treatment of dark energy and dark matter. Also, a better understanding of how, within the discussed scenario, non-gravitational interactions could be suppressed in cosmological models, is another interesting issue deserving consideration.

\section*{Acknowledgements}
M.K. has been supported by a Juan de la Cierva-incorporación grant (IJC2020-042690-I). This work has been partially supported by MICINN (Spain), project PID2019-104397GB-I00, of the Spanish State Research Agency program AEI/10.13039/501100011033, by the Catalan Government, AGAUR project 2021-SGR-00171, and by the program Unidad de Excelencia María de Maeztu CEX2020-001058-M.


\begin{thebibliography}{1}

\bibitem{Aghanim_H0}
N. Aghanim et al. (Planck), Astron. Astrophys. 641, A6 (2020), 1807.06209.

\bibitem{Riess_H0}
A. G. Riess et al., Astrophys. J. 861, 126 (2018), 1804.10655.

\bibitem{Wong_H0}
K. C. Wong et al, Mon. Not. Roy. Astron. Soc. 498, 1420 (2020), 1907.04869.

\bibitem{Freedman_H0}
W. L. Freedman et al, (2019), 1907.05922.

\bibitem{Perivolaropoulos_LCDM}
L. Perivolaropoulos, F. Skara, New Astronomy Reviews, Volume 95, 2022, 101659.

\bibitem{Elizalde_H0}
E. Elizalde, J. Gluza, M. Khurshudyan, arXiv:2104.01077.

\bibitem{Elizalde_H0_1}
M. Khurshudyan, Astrophysics 66 (3), 423-440, 2023.

\bibitem{Elizalde_0}
Y. F. Cai et al, Astrophys.J. 888, 62 (2020).

\bibitem{Elizalde_1}
E. Elizalde, M. Khurshudyan, Phys.Rev.D 99 (2019) 10, 103533.

\bibitem{Elizalde_2}
E. Elizalde, M. Khurshudyan, Eur. Phys. J. C 82 (2022) 9, 81.

\bibitem{Elizalde_3}
E. Elizalde et al, arxiv:2203.06767.

\bibitem{Elizalde_4}
M. Aljaf et al, Eur. Phys. J. C 82 (2022) 12, 1130.

\bibitem{Elizalde_5}
K. Yerzhanov et al,  Mod. Phys. Lett. A 36 (2021) 31, 2150222.

\bibitem{Elizalde_6}
M. Aljaf et al, Int. J. Mod. Phys. A 37 (2022) 34, 2250211.

\bibitem{Elizalde_7}
E. Elizalde, M. Khurshudyan, Eur.Phys.J.C 81 (2021) 4, 335, Eur.Phys.J.C 81 (2021) 5, 438 (erratum).

\bibitem{Elizalde_8}
E. Elizalde, M. Khurshudyan, Phys. Dark Univ. 37 (2022) 101114.

\bibitem{Elizalde_9}
E. Elizalde et al, Phys.Rev.D 102 (2020) 12, 123501.

\bibitem{MK1}
E. Elizalde et al, Int. J. Mod. Phys. D 28, No. 01, 1950019 (2019).

\bibitem{MK13}
M. Khurshudyan, R. Myrzakulov, Eur. Phys. J. C 77: 65 (2017).  

\bibitem{MK12}
C. Li et al, Phys. Lett. B 80, 135141(2020).

\bibitem{Odintsov:2017icc}
 S. D. Odintsov et al., Phys. Rev. D 96, no.4, 044022 (2017).
 
 \bibitem{Bamba:2012cp}
K. Bamba et al, Astrophys. Space Sci. 342, 155 (2012). 
 
\bibitem{MK8}
W. Yang et al, JCAP 1911, 044  (2019). 


\bibitem{Mk14}
W. Yang et al, Phys. Rev. D 99, no.4, 043543 (2019). 

\bibitem{Mk15}
E. Elizalde, M. Khurshudyan, Int. J. Mod. Phys. D 27, No. 04, 1850037 (2018).

\bibitem{Mk17}
M. Aljaf et al, Eur. Phys. J. C  80:112 (2020).

\bibitem{Mk18}
E. Sadri et al, Eur. Phys. J. C  80:393 (2020).

\bibitem{H0start_1}
H. Amirhashchi, A. K. Yadav, arxiv:2001.03775.

\bibitem{H0start_4}
G. Alestas et al, arxiv:2004.08363.

\bibitem{H0start_5}
D. Wang, D. Mota, arxiv:2003.10095.

\bibitem{H0start_6}
N. Blinov et al, arxiv: 2004.06114.

\bibitem{H0start_7}
E.K. Li et al, arxiv: 1911.12076.

\bibitem{H0start_10}
E. Di Valentino et al, Phys. Rev. D 101, 063502 (2020).

\bibitem{H0start_12}
E. Di Valentino et al, arxiv: 2005.12587.

\bibitem{H0End}
R. C. Nunes, JCAP 05, 052 (2018).

\bibitem{INStart}
I. Brevik et al, Int. J. Geom. Meth. Mod. Phys. 14, 1750185 (2017).

\bibitem{INStart_2}
S. Capozziello et al, Phys.Rev. D99, 023532  (2019).

\bibitem{INStart_4}
S. Nojiri, S. D. Odintsov, Phys. Rev. D 72, 023003 (2005).

\bibitem{INStart_5}
I Brevik et al, Mod. Phys. Lett. A 27, 1250210 (2012).

\bibitem{INStart_7}
I. Brevik et al, Int. J. Mod. Phys. D 26, no.14, 1730024 (2017).

\bibitem{INStart_9}
S. D. Odintsov et al, Annals Phys. 398, 238-253  (2018).

\bibitem{INEnd}
S. D. Odintsov et al,  Phys. Rev. D 101, 044010 (2020).

\bibitem{GP_0}
X. Rin et al, Astrophys. J. 932 (2022) 2, 131.

\bibitem{GP_1}
Peng-Ju Wu et al, arxiv:2209.08502.

\bibitem{GP_2}
J. L. Said et al, JCAP 06, 015  (2021).

\bibitem{GP_3}
A. Gomez-Valent, Luca Amendola,  JCAP, 1804, 051 (2018).

\bibitem{GP_4}
S. Dhawan et al, Mon. Not. Roy. Astron. Soc. 506, L1 (2021).

\bibitem{GP_5}
E. O. Colgain, M. M. Sheikh-Jabbari, arXiv:2101.08565.

\bibitem{GP_6}
R. C. Bernardo, J. L. Said, JCAP 09, 014  (2021).

\bibitem{GP_7}
R. C. Bernardo, J. L. Said,  JCAP 08, 027 (2021).

\bibitem{HTable_0}
C. Zhang et al, Research in Astronomy and Astrophysics 14, 1221 (2014). 

\bibitem{HTable_1}
M. Moresco et al, JCAP 05, 014 (2016).

\bibitem{HTable_2}
R. Jimenez et al,  Astrophys. J. 593, 622 (2003).

\bibitem{HTable_3}
D. Stern et al, JCAP 008 (2010).

\bibitem{HTable_4}
M. Moresco et al, JCAP 08, 006 (2012).

\bibitem{HTable_5}
J. Simon et al, Phys. Rev. D 71, 123001 (2005).

\bibitem{HTable_6}
M. Moresco, Mon. Not. R. Astron. Soc. Lett. 450, L16 (2015).

\bibitem{HTable_7}
E. Gaztanaga et al, Mon. Not. R. Astron. Soc. 399, 1663 (2009).

\bibitem{HTable_8}
C. Blake et al, Mon. Not. R. Astron. Soc. 425, 405 (2012).

\bibitem{HTable_9}
X. Xu et al, Mon. Not. R. Astron. Soc. 431, 2834 (2013).

\bibitem{HTable_10}
T. Delubac et al, Astronomy $\&$ Astrophysics 552, A96 (2013).

\bibitem{HTable_11}
T. Delubac et al, Astronomy $\&$ Astrophysics 574, A59 (2015).

\bibitem{HTable_12}
L. Samushia et al, Mon. Not. R. Astron. Soc. 429, 1514 (2013).

\bibitem{HTable_13}
A. Font-Ribera et al, JCAP 05, 027 (2014).

\bibitem{Seikel}
M. Seikel, C. Clarkson and M. Smith, JCAP 06, 036 (2012).

















 












\end{thebibliography}
\end{document}